\documentclass[aps,prx,twocolumn]{revtex4-1}
\usepackage[utf8]{inputenc}
\usepackage[english]{babel}
\usepackage[T1]{fontenc}
\usepackage{lmodern}
\usepackage{graphicx}
\usepackage{amsmath}
\usepackage{amsfonts}
\usepackage{amssymb}
\usepackage{microtype}
\usepackage{braket}
\usepackage{subfigure}
\usepackage{etoolbox}
\usepackage[breaklinks=true,colorlinks,citecolor=blue]{hyperref}

\newcommand{\beq}{\begin{eqnarray}}
	\newcommand{\eeq}{\end{eqnarray}}
\newcommand{\be}{\begin{equation}}
	\newcommand{\ee}{\end{equation}}

\begin{document}

\preprint{APS/123-QED}

\title{Dynamical crossover behaviour in the relaxation  of  quenched quantum many-body systems}

\author{Aamir Ahmad Makki}
\author{Souvik Bandyopadhyay}
\author{Somnath Maity}

\author{Amit Dutta}
\affiliation{Department of Physics, Indian Institute of Technology Kanpur, Kanpur 208016, India}



\begin{abstract}
A crossover between different power-law relaxation behaviors of many-body periodically driven integrable systems has come to light in recent years. We demonstrate using integrable quantum systems, that similar kinds of dynamical transitions may also occur in the relaxation of such systems following a sudden quench. Particularly, we observe two distinct power-law relaxation behaviors following a sudden quench in the integrable XY model, depending upon whether the quenched Hamiltonian lies in the commensurate or the incommensurate phase. The relaxation behavior for quenches at and near the boundary line, called the
disorder line (DL), separating these phases is also characterized. The relaxation at the DL shows a new scaling exponent previously unexplored. The transitions occur through a crossover from the commensurate/incommensurate scaling behavior to the DL scaling behavior. The crossover time diverges like a power law as the parameters of the final quenched Hamiltonian approach the DL. The transitions are also observed to be robust under weak integrability breaking perturbations but disappear following strongly chaotic quenches.
\end{abstract}

\maketitle


\section{\label{sec:level1}Introduction}

Phase transitions, both classical and quantum, have been studied extensively \cite{goldenfeld,sachdev_2011,polkovnikovrev,dziarmaga,ADbook,mondal2010}. However, over the past decade, the notion of phase transition has been extended to the dynamics of non-equilibrium quantum many-body systems. In a seminal work, Heyl \textit{et al.,} first showed \cite{Heyl} the existence of a {dynamical} quantum phase transition (DQPT) manifested as non-analyticities in the temporal evolution of the Loschmidt overlap following a quantum quench (For a review see \cite{Heylreview}). In parallel, Sen \textit{et al.,} \cite{sen2016,Nandy_2018} established the existence of another class of dynamical phase transitions in periodically driven integrable systems in terms of the relaxation behavior of two-point correlators.
The difference of the correlators at a time $t$ from their respective steady state values,  reached asymptotically, scales as a power law in time, $t^{-\mu}$. For drive frequencies $\omega$ greater than a critical frequency, $\omega_c$, the exponent $\mu=(d+2)/2$, where $d$ is the dimensionality of the system. As $\omega$ is decreased to $\omega<\omega_c$, $\mu$ changes to $d/2$. This transition has been argued to be an artefact of a qualitative change in the Floquet quasienergy spectrum.  More precisely, it depends on whether the zeros of the group velocity all lie on the Brillouin zone edges and center. This is because the time dependent part of the correlation functions reduces to a saddle point integral where the saddles are located at the zeros of the group velocity. For $\omega<\omega_c$, there is a zero of the group velocity lying inside (not at edges or center) the Brillouin zone and the contribution from this extra saddle gives rise to a different scaling with $\mu=d/2$. 

In this paper, we probe the existence of the second type of transitions in quenched quantum systems where the subsequent time evolution is determined by the final Hamiltonian. To this end we will consider two different integrable models: the 1D nearest neighbour spin-$1/2$ XY model \cite{LIEB,bunder,PFEUTY,mbeng} and an extended Ising model which includes three spin interactions \cite{zhangprl,divakaran,sourav,jafari}. The motivation for considering these models is that they show similar changes in the energy spectrum in terms of the location of the zeros of the group velocity. On the other hand, the study of thermalization of closed quantum systems thrown out of equilibrium has become essential to understand the emergence of quantum statistical mechanics. Despite the presence of an extensive number of conserved quantities, integrable quantum many body systems in the thermodynamic limit are known to reach a statistical description after sufficiently long time evolutions. Namely, the generalised Gibbs ensemble (GGE) is completely defined by the extensive number of conserved quantities, which act as constraints preventing an ergodic relaxation into a thermal ensemble \cite{Rigol,Nandy2016}. On the contrary, short range interacting strongly nonintegrable many body systems, lacking sufficient number of conserved quantities to prevent thermalization, are known to satisfy the eigenstate thermalization hypothesis (ETH) \cite{Rigol}. In thermodynamically large systems, the ETH directly implies a microcanonical description of the late time statistics of local observables. However, the late time dynamics leading to the eventual thermalization and the relaxation time scales involved, in both integrable and chaotic many body systems is yet to be understood exhaustively. There is a recent upsurge in unraveling the signatures of quantum criticality in the late time statistical behavior of local observables in such systems \cite{dag2021probing,haldar}. In this light, it raises the inevitable question of whether the eventual relaxation of chaotic many body systems also show robust signatures of the dynamical phase transitions and the crossover time scales. To further the generic nature of the dynamical relaxation crossovers, we probe their robustness against weak integrability breaking perturbations. Remarakably, we establish that in finite size systems the dynamical transitions are indeed persistent following weakly chaotic quenches. However, following generic strongly chaotic quenches we observe the disappearance of all slow power-law relaxation of local observables. Rather, we observe a fast exponential decay of correlations following strongly chaotic quenches. We note that an exponential decay of correlations has also been observed in a periodically driven integrable system coupled to a bath \cite{PhysRevB.102.235154}.\\

In Sec.~\ref{sec2}, we present the XY model Hamiltonian and its spectrum, phase diagram, ground state and time evolution following a quench. Depending on the parameters  of the final Hamiltonian (i.e., its  position on the phase diagram of the model),  we find two different power law scalings in the decay of two-point correlation functions to their respective steady states. In Sec.~\ref{sec3}, we investigate how the transition between the different power laws occurs and what happens at the phase boundary. We find that at the phase boundary, the exponent of the power law scaling is different than in either of the phases. Moreover, close to the boundary, the scaling remains that of the boundary for smaller times until a crossover to the scaling behavior of the corresponding phase. The crossover time is shown to diverge upon approach to the boundary. The divergence takes the form of a power law in the relevant parameter of the Hamiltonian. In Sec.~\ref{sec:extended_ising}, we discuss the dynamical phase transitions in the extended Ising model. We also show a different kind of crossover than the one discussed in Sec.~\ref{sec3}. This crossover is between the two scaling exponents, from $\mu=3/2$ to $\mu=1/2$ and  the crossover time in this case does not show any divergence. This crossover occurs due to relative size of contributions from different saddle points to the saddle point integral. In Sec.~\ref{sec5}, we  revisit the XY model in light of the latter kind of crossover. We also note the competition between different crossovers. In Sec.~\ref{sec6}, we show that the transitions between the scalings for the integrable case persist for finite size systems in the presence of weak integrability breaking perturbations. Concluding comments are
presented in Sec.~\ref{sec_conclusion}.  In Appendix~\ref{appendix_1}, we elucidate an assumption in the argument for calculating the analytical value of the crossover exponent. In Appendix~\ref{appendix_2}, we provide exact diagonalization results for the integrable case.

\section{XY Model}\label{sec2} 

\subsection{Spectrum and the Ground State}
We consider the following spin Hamiltonian \cite{mbeng,PFEUTY,bunder} on a 1D lattice with unit spacing and total length $L$:
\begin{eqnarray}
H_{XY}=-\sum_{i=1}^L \left(\frac{1+\chi}{2}\right)\sigma_{i}^x\sigma_{i+1}^x+\left(\frac{1-\chi}{2}\right)\sigma_{i}^y\sigma_{i+1}^y\nonumber\\
-\sum_{i=1}^Lh\sigma_i^z\textnormal{,\quad}
\label{XYHam}\end{eqnarray}
where $\sigma_i^{\alpha},(i=1,...,L;\alpha=x,y,z)$ are the Pauli spin matrices and $h$ is an applied magnetic field. $\chi$ is an anistropy parameter which under the limits $\chi\xrightarrow{}0$ and $\chi\xrightarrow{}1$ reduces to the isotropic XY model and the transverse field ising model (TFIM) respectively. For the sake of simplicity, we will be considering $L$ to be even.\\

The Hamiltonian in Eq.~\ref{XYHam} is readily diagonalizable \cite{PFEUTY} by invoking upon a Jordan-Wigner (JW) transformation \cite{LIEB} to spinless fermionic operators $c_i,c_i^{\dagger}$. Using the JW transformation, the Hamiltonian can be rewritten as:
\begin{eqnarray}
H_{XY}=-\sum_{i=1}^L \left(c_j^{\dagger}c_{j+1}+\chi c_j^{\dagger}c_{j+1}^{\dagger}+\textrm{H.c.}\right)\nonumber\\
+h\sum_{i=1}^L\left(2c_j^{\dagger}c_j-1\right),
\label{XYFerm}\end{eqnarray}
where $\textrm{h.c.}$ denotes Hermitian conjugate. We will stick with the even fermionic number parity sector with anti-periodic boundary conditions (ABC), $$c_{L+1}=-c_1.$$

\begin{figure*}
\subfigure[]{
	\includegraphics[width=6.0cm,height=5.0cm]{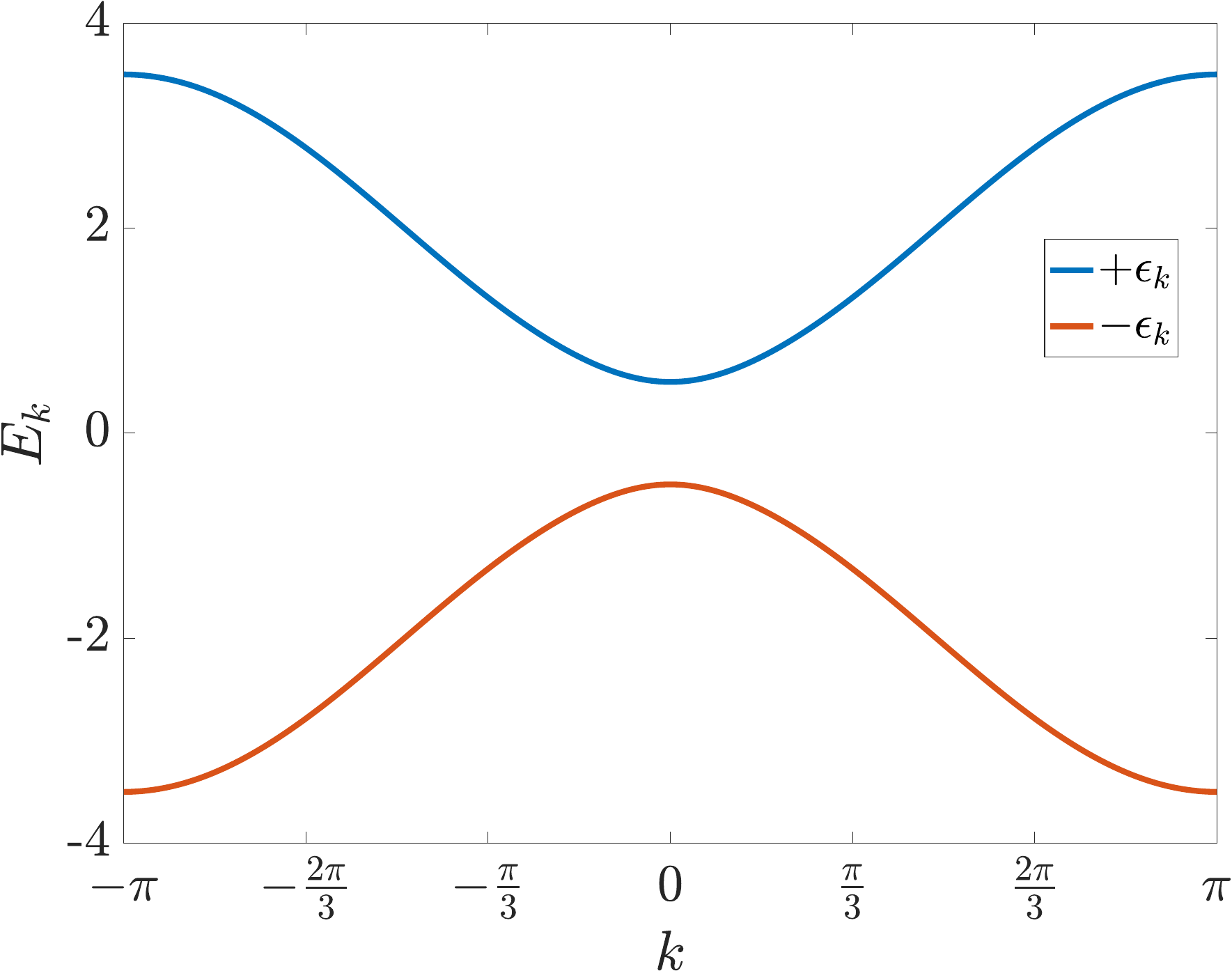}
	\label{f1}}
\hspace{1.5cm}
\subfigure[]{
	\includegraphics[width=6.0cm,height=5.0cm]{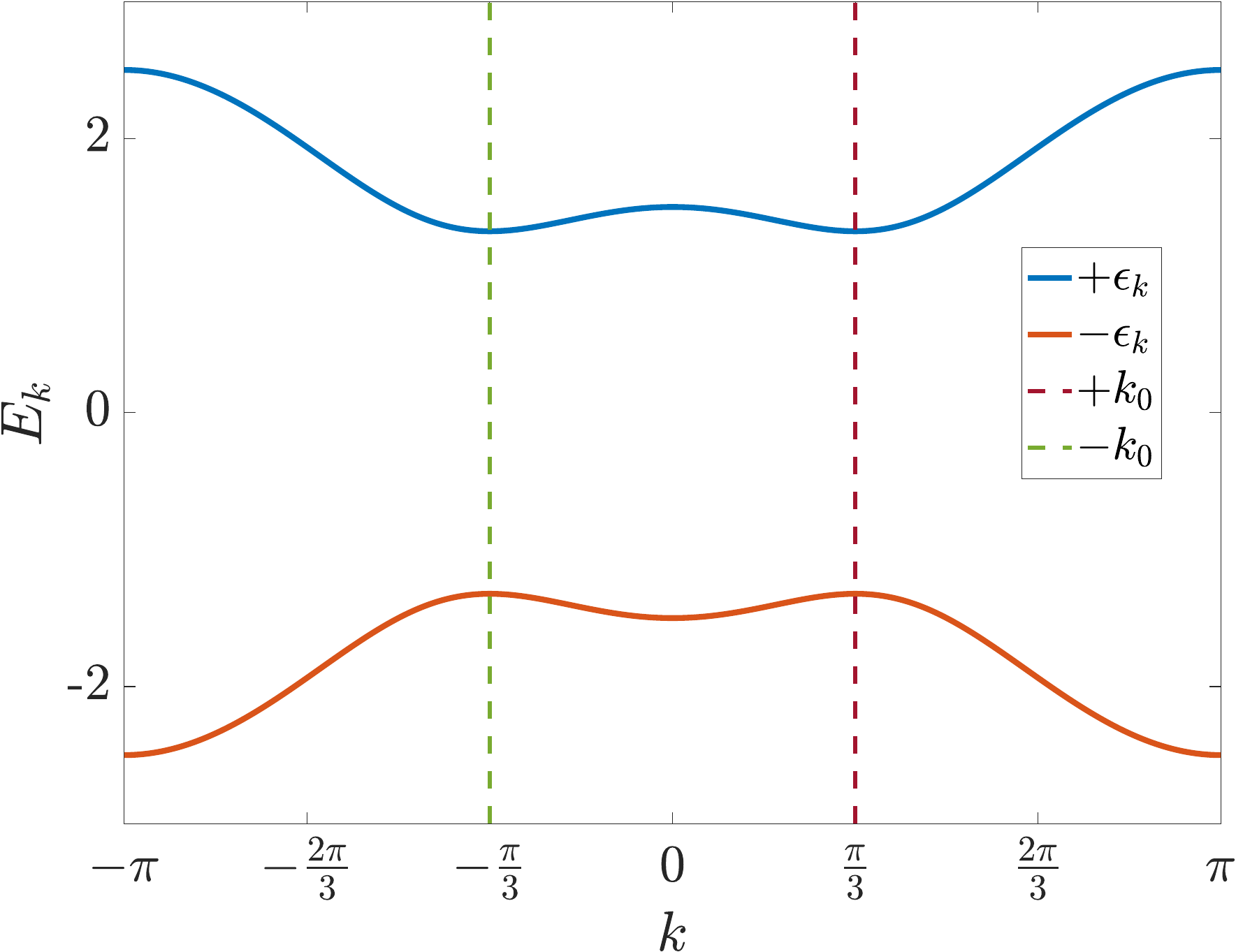}
	\label{f4}}
\caption{The energy spectrum for the commensurate and incommensurate phases. (a) Commensurate phase (with $h=0.75$, $\chi=\frac{1}{\sqrt{2}}$):  the extrema of $\epsilon_k$ all lie at the edges and center of the Brillouin zone. (b) Incommensurate phase (with $h=0.25$, $\chi=\frac{1}{\sqrt{2}}$.): the minimum for the positive energy band occurs at $k_0\neq 0,\pm\pi$.}
\end{figure*}

Taking a Fourier transform of the fermionic operators, we obtain the momentum space representation of the Hamiltonian,
\begin{eqnarray}
H_{XY}=\displaystyle{\sum\limits_{k>0}H_{k}}\nonumber=\sum\limits_{k}\mathbf{\Psi}_k^\dagger\mathbf{H}_k\mathbf{\Psi}_k \quad\quad\quad\quad\nonumber\\=\sum_k
\begin{pmatrix}
c_k^{\dagger}&c_{-k}
\end{pmatrix}\underset{\mathbf{H}_k}{\underbrace{
\begin{pmatrix}
2(h-\cos{k})&-i2\chi\sin{k}\\
i2\chi\sin{k}&-2(h-\cos{k})
\end{pmatrix}}}
\begin{pmatrix}
c_k\\
c_{-k}^{\dagger}
\end{pmatrix},\nonumber\\
\label{Hkmat}\end{eqnarray}
where $\mathbf{\Psi}_k=(
c_k,c_{-k}^{\dagger}
)^{T}$ are Nambu spinors.  {The anti-periodic boundary conditions require that the wavevectors $k$ belong to the set
\[\mathcal{K}=\left\{\pm\frac{(2n-1)\pi}{L},n=1,2,\dots,\frac{L}{2}\right\}.\]} We note, that the matrix $\mathbf{H}_k$ has the form of a Hamiltonian resembling a single spin in a magnetic field, i.e., it can be written as
\begin{eqnarray}
\mathbf{H}_k=\mathbf{R}_k\cdot\mathbf{\tau},
\label{Hkspin}\end{eqnarray}
where $\mathbf{\tau}$ is a vector of Pauli matrices and $\mathbf{R}_k=\begin{pmatrix}
x_k,y_k,z_k
\end{pmatrix}^{T}=2\begin{pmatrix}
0,\chi\sin{k},h-\cos{k}
\end{pmatrix}^{T}$. The eigenenergies can then be obtained straightforwardly to be $E_k=\pm\epsilon_k$ \cite{mbeng}, where,
\begin{eqnarray}
\epsilon_k=|\mathbf{R}_k|=2\sqrt{(h-\cos k)^2+\chi^2\sin^2k}.\label{dispersion}\end{eqnarray}
\begin{figure}[t]
\includegraphics[width=6cm, height=5cm]{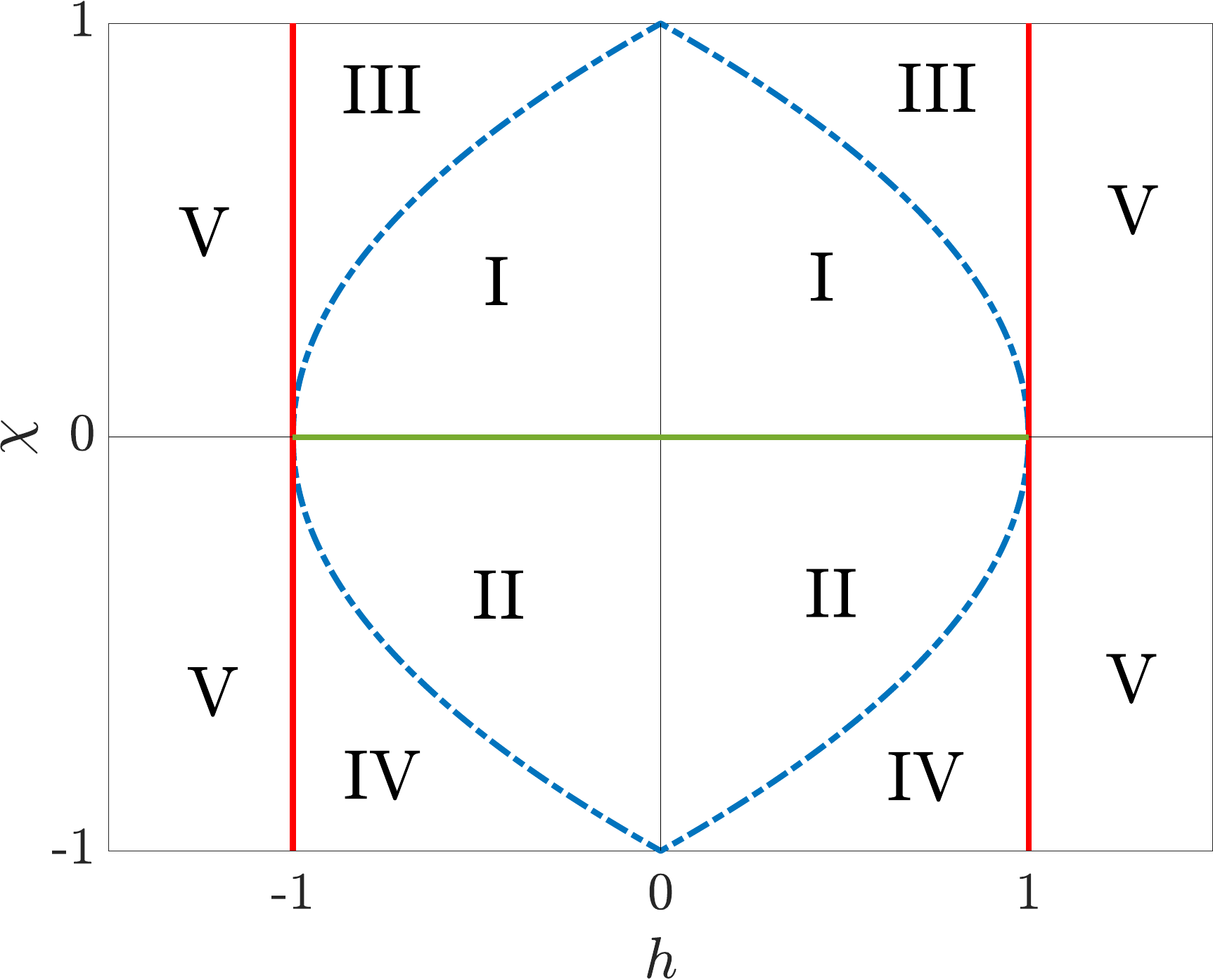}
\caption{\label{XYPD}Phase diagram of the XY model in \eqref{XYHam}.   {The red, blue and
green lines mark the phase boundaries in the equilibrium model. The red and green lines mark quantum phase transitions while the blue broken line marks a crossover.}  The red line represents the ferromagnetic to paramagnetic transition. On the other hand,  the blue line, known as the disorder line (DL) and described by the by the equation $|h|=|1-\chi^2|$, separates the  commensurate and incommensurate phases.  Finally, the green line denotes the anisotropic XY transition between two ferromagnetic phases. The associated phases are I: incommensurate, ferromagnetic along X; II: incommensurate, ferromagnetic along Y; III: commensurate, ferromagnetic along X; IV: commensurate, ferromagnetic along Y; V: paramagnetic, commensurate}
\end{figure}
The various phases exhibited by the XY model can be compiled into a phase diagram \cite{ADbook} (Fig.~\ref{XYPD}). The system has quantum critical points (QCP) at $h=\pm1$ where the gap between the positive and negative energy bands of the spectrum closes. For the $h=1$ QCP, the gap closing occurs at $k=0$, while for $h=-1$, it occurs at $k=\pm\pi$. The model exhibits a quantum phase transition (QPT) from a ferromagnetic (regions I-IV) to a paramagnetic (region V) phase as shown in Fig.~\ref{XYPD}. The ferromagnetic phase occurs for magnetic fields $|h|<1$ while the paramagnetic phase occurs for $|h|>1$. For $0<\chi\leq1$ (regions I and III), the order parameter for the ferromagnetic phase is $\langle\sigma^x\rangle$ while for $-1\leq\chi<0$ (regions II and IV) the order parameter is $\langle\sigma^y\rangle$. We call this transition the anisotropic X-Y transition (green line in Fig.~\ref{XYPD}). Note that we are not considering frustrated systems where $|\chi|>1$. 
There is another type of transition, namely the commensurate-incommensurate transition \cite{bunder}. In the commensurate phase (regions III-V), the minimum band gap lies at the extremities of the Brillouin zone while in the incommensurate phase (regions I and II), the band gap lies within the Brillouin zone (Figs.~\ref{f1},\ref{f4}). To see this analytically, consider the derivative of $\epsilon_k$ with respect to the wavevector $k$: $d\epsilon_k/dk$ always has zeros at $k=0,\pm\pi$. But, for $|h|<|1-\chi^2|$, there is an additional zero of $d\epsilon_k/dk$ at $k_0$ such that,
\[\cos{k_0}=\frac{h}{1-\chi^2}.\]
The boundary between the commensurate and the incommensurate phase (blue line in Fig.~\ref{XYPD}) is known as the disorder line (DL).   {In this work, we will focus mainly on the difference between quenches to regions on either side of this line, starting from the ground state at $h\rightarrow{}\infty$.}\\

In order to find the ground state, we need to consider the eigenvectors of $\mathbf{H}_k$.  {Suppose the positive energy eigenvector of $\mathbf{H}_k$ is given by $(u_k,v_k)^T$, then the negative energy eigenvector is given by $(-v_k^*,u_k^*)^T$. Also,  as is evident from Eq.~\eqref{Hkspin}, since $\mathbf{R}_k$ lies in the $y-z$ plane, $u_k$ is real and $v_k$ is purely imaginary in this case. The quasiparticle excitations of the system are given by linear combinations of JW fermions,
\begin{eqnarray}
\gamma_k=u_k^*c_k+v_k^*c_{-k}^{\dagger},\textnormal{  }\gamma_{-k}^{\dagger}=-v_kc_k+u_kc_{-k}^{\dagger}.
\end{eqnarray}
}The ground state, $|\phi\rangle$ is determined by the condition that it is annihilated by all of the $\gamma_k$'s, that is $\gamma_k|\phi\rangle=0\textnormal{ }\forall\textnormal{ } k$. This condition gives 
\begin{equation}
|\phi\rangle=\prod_{k>0,k\in\mathcal{K}}\left(u_k+v_kc_k^{\dagger}c_{-k}^{\dagger}\right)|0\rangle
.\end{equation}
We note that this is similar to the so called Bardeen-Cooper-Schrieffer (BCS) ground state of a superconductor.

\subsection{Quench dynamics}
\subsubsection{Bogoliubov-de Gennes Equations}
Consider, starting from an initial state $\ket{\psi(0)}$ and subjecting the system to a sudden quench at time $t=0$. The system is then allowed to evolve freely with the quenched Hamiltonian $H$.
To obtain a solution of the time dependent Schrödinger equation,
\[i\partial_t|\psi(t)\rangle=H|\psi(t)\rangle,\]
we make the ansatz that the state is in a BCS type state at all times:
\begin{equation}
|\psi(t)\rangle=\prod_{k>0,k\in\mathcal{K}}\left(u_k(t)+v_k(t)c_k^{\dagger}c_{-k}^{\dagger}\right)|0\rangle,
\label{ansatz}\end{equation}
where $u_k(t)$ and $v_k(t)$ are time dependent coefficients.
Plugging this ansatz into the Schrödinger equation and matching coefficients of the orthogonal states, we get the BdG equations,
\begin{equation}
i\partial_t\mathbf{\psi}_k(t)=\mathbf{H}_k\mathbf{\psi}_k(t),
\label{BdG}\end{equation}
where $\mathbf{\psi}_k=\left(v_k(t),u_k(t)\right)^T$. Eq.~\ref{BdG} has the formal solution
\begin{equation}
\mathbf{\psi}_k(t)=e^{-i\mathbf{H}_kt}\mathbf{\psi}_k(0).
\label{solBdG}\end{equation}
Because of the simple form of the single-particle Hamiltonian, $\mathbf{H}_k$, the matrix exponential can be easily calculated and we get the exact solutions to the BdG equations,
\begin{subequations}
\label{vkuk}
\begin{eqnarray}
&v_k(t)=v_k(0)\big[\cos{(\epsilon_kt)}-i\sin{\alpha_k}\sin{(\epsilon_kt)}\big]\nonumber\\&\textnormal{\quad\quad\quad\quad}-u_k(0)\big[\cos{\alpha_k}\sin{(\epsilon_kt)}\big]\label{vk},\\
&u_k(t)=u_k(0)\big[\cos{(\epsilon_kt)}+i\sin{\alpha_k}\sin{(\epsilon_kt)}\big]\nonumber\\&\textnormal{\quad\quad\quad\quad}+v_k(0)\big[\cos{\alpha_k}\sin{(\epsilon_kt)}\big],\label{uk}
\end{eqnarray}
\end{subequations}
where we have defined $\cos{\alpha_k}=\frac{y_k}{\epsilon_k}$ so that $\sin{\alpha_k}=\frac{z_k}{\epsilon_k}$.
\subsubsection{Correlation Functions}
To observe the dynamical transition in the relaxation behavior following the quench, we require the fermionic two point correlation functions $C_{mn}(t)=\langle c_m^{\dagger}c_n\rangle_t$ and $F_{mn}(t)=\langle c_m^{\dagger}c_n^{\dagger}\rangle_t$ where $\langle.\rangle_t$ indicates that the expectation value is taken over the state $|\psi(t)\rangle$ as defined in Eq.~\ref{ansatz}. By transforming the fermionic operators to the fourier space we get the following expressions for $C_{mn}(t)$ and $F_{mn}(t)$,
\begin{subequations}
\label{corr}
\begin{eqnarray}
C_{mn}(t)=\frac{2}{L}\sum_{k>0,k\in\mathcal{K}}|v_k(t)|^2\cos{k(m-n)},\\
F_{mn}(t)=\frac{2}{L}\sum_{k>0,k\in\mathcal{K}}v^*_k(t)u_k(t)\sin{k(m-n)}.
\end{eqnarray}
\end{subequations}
All other two-point correlations can be obtained from these two by complex conjugation and/or exchanging indices. We will focus on the single particle ($C_{mn}$) correlations from here on. Similar results follow for the pair creation ($F_{mn}$) terms.
\subsection{Post-quench Dynamics and Dynamical Phase Transition}
We demonstrate the dynamical phase transition using an initial state in which all single particle correlations vanish. Such a state is the fermionic vacuum or equivalently in the spin picture, the state $|\uparrow\uparrow\uparrow\dots\rangle$. This state is the ground state of $H_{XY}$ for $h\xrightarrow{}\infty$. Thus, starting from the ground state of the $h\xrightarrow{}\infty$ Hamiltonian, we quench it to $h=h_f$ at $t=0$. The initial state $\mathbf{\psi}_k(0)$ is given by
$\mathbf{\psi}_k(0)=
\begin{pmatrix}
v_k(0),u_k(0)
\end{pmatrix}^T=\begin{pmatrix}
0,1
\end{pmatrix}^T\forall k.
$\\

For such an initial state, the time dependent correlation functions are (in the $L\xrightarrow{}\infty$ limit),
\begin{eqnarray*}
&\displaystyle{C_{mn}(t)=\frac{1}{\pi}\int_0^{\pi}dk|v_k(t)|^2\cos{[k(m-n)]}}\\
&\displaystyle{=\frac{1}{\pi}\int_0^{\pi}dk\cos^2\alpha_k\sin^2{(\epsilon_kt)}\cos{[k(m-n)]}}{\quad\quad}\\&=\underset{\textnormal{Steady State: }C_{mn}(\infty)}{\underbrace{\frac{1}{2\pi}\int_0^{\pi}dk\cos^2\alpha_k\cos{[k(m-n)]}}}\textnormal{\quad\quad\quad\quad\quad\quad}\\&\textnormal{\quad\quad}\displaystyle{\underset{-\delta C_{mn}(t)}{\underbrace{\displaystyle{-\frac{1}{2\pi}\int_0^{\pi}dk\cos^2\alpha_k\cos{(2\epsilon_kt)}\cos{[k(m-n)]}}}}}.
\end{eqnarray*}
We proceed by rewriting as $\delta C_{mn}(t)=I(t)+I(-t)$, where
\begin{equation}
I(t)= \frac{1}{4\pi}\int_0^{\pi}dk\cos^2\alpha_ke^{i2\epsilon_kt}\cos{[k(m-n)]}.
\end{equation}

\begin{figure*}
\subfigure[]{
	\includegraphics[width=7.0cm,height=5.0cm]{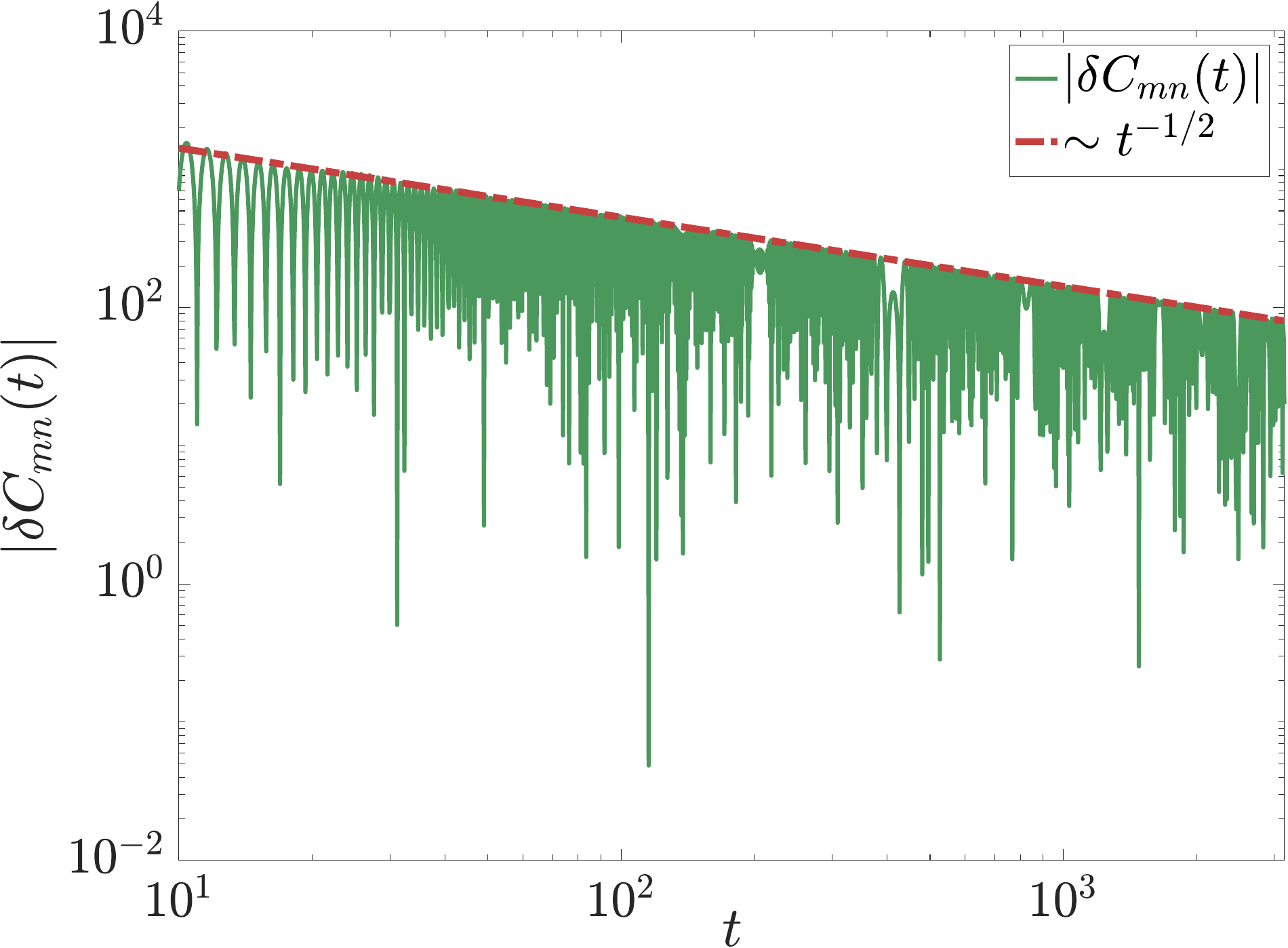}
	\label{incomm}}
\hspace{0.5cm}
\subfigure[]{
	\includegraphics[width=7.0cm,height=5.0cm]{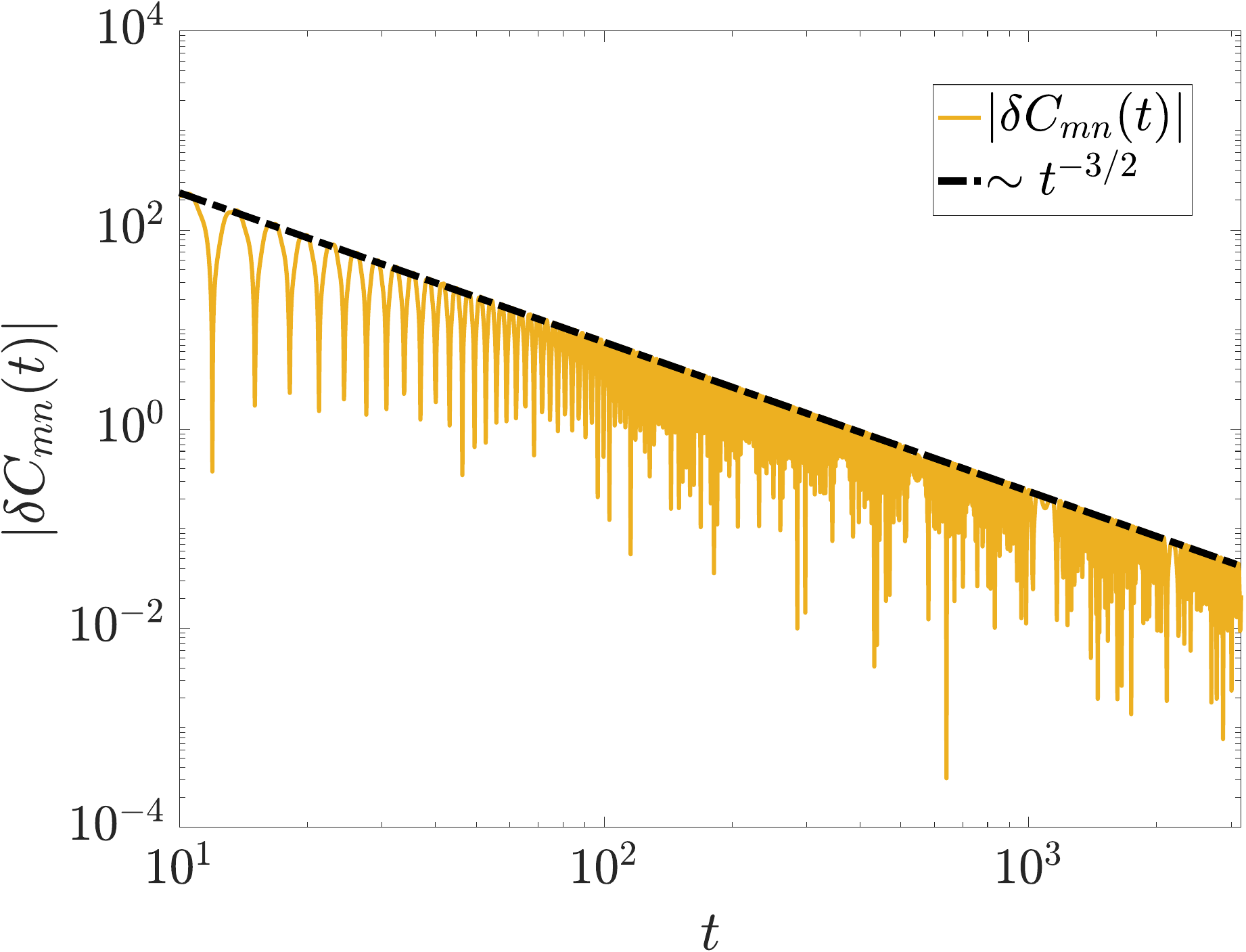}
	\label{comm}}
\caption{Dynamical phase transition when the Hamiltonian Eq.~\eqref{XYHam}  is quenched to the  (a) incommensurate phase for parameters $h=0.25$, $\chi=\frac{1}{\sqrt{2}}$ and (b) to the commensurate phase with parameters $h=0.75$, $\chi=\frac{1}{\sqrt{2}}$. We observe distinct power-law decays of the correlation $C_{mn}$ with different exponents in two cases. In both the cases $|n-m|=1$. The correlation function has been scaled up by a factor of $L=40000$ to avoid numerical errors due to numbers below machine precision.  {The prefactors for the power law are calculated in Appendix \ref{appendix_0}.}}	
\end{figure*}

Now we have two cases depending on whether $h_f$ lies in the commensurate or the incommensurate phase.
\subsubsection{Quench to the incommensurate phase}
In this section we deal with the situation in which the quenched Hamiltonian lies in region I of the phase diagram (Fig.~\ref{XYPD}). To proceed, we note that at large times, the exponential term in $I(t)$ oscillates rapidly and so we can make the stationary phase approximation (SPA), that is, the integral is dominated by the contributions near the extrema of $\epsilon_k$, where we can replace $e^{i2\epsilon_kt}$ by a gaussian in the variable $k$. More precisely, 
\begin{equation}\label{eq:SPAincomm}
I(t)\approx \mathcal{C}e^{i2\epsilon_{k_0}t}\int_{-\infty}^{\infty}dk\exp\Bigg[{i\frac{d^2\epsilon_k}{dk^2}\Bigg|_{k_0}(k-k_0)^2t}\Bigg],
\end{equation}
where $\mathcal{C}=\cos^2\alpha_{k_0}\cos{[k_0(m-n)]}/4\pi$. By performing the gaussian integral, we get the scaling relation,
\begin{equation}
\delta C_{mn}(t)\sim t^{-1/2}.
\end{equation}
Results of the numerical simulation (Fig.~\ref{incomm}) agree with this scaling.  {For exact expressions for the coefficients of the power laws, see Appendix \ref{appendix_0}.}
\subsubsection{Quench to the commensurate phase}
In this part, the Hamiltonian is quenched to the phase III of the phase diagram in Fig.~\ref{XYPD}. In this case  more care is needed in applying the SPA because the extrema of $\epsilon_k$ lie at the Brillouin zone edges and center, where $\cos^2\alpha_k=\frac{\chi^2\sin^2k}{\epsilon_k^2}$ vanishes, so we cannot simply scale it out of the integral. To proceed, we take the leading order (in $k$)  contribution near the saddle point, $\cos^2\alpha_k\approx\frac{\chi^2}{(h-1)^2}k^2$.
Thus,
\begin{equation}
    I(t)\sim\int_{-\infty}^{\infty}dkk^2\exp\Bigg[{i\frac{d^2\epsilon_k}{dk^2}\Bigg|_{0}k^2t}\Bigg].
\end{equation}
This integral can again be performed readily, upon which we get the scaling relation,
\begin{equation}
\delta C_{mn}(t)\sim t^{-3/2}.
\end{equation}
This result also agrees with the numerical simulations (Fig.~\ref{comm}).
From the above change in scaling, we see that there is a transition in the post-quench relaxation behavior of the correlation functions depending on whether we quench the system to the commensurate or the incommensurate phase.\\

To get a physical picture of the mechanism behind the transition, we provide here a heuristic argument. Consider the expression for the time dependent part of the correlation function $C_{mn}$, {
\[\delta C_{mn}(t)\sim\int dk \cos^2{\alpha_k}e^{i2\epsilon_kt}e^{ik(m-n)},\]
}where we have rewritten $\cos{k(m-n)}$ as $[e^{ik(m-n)}+e^{-ik(m-n)}]/2$ and kept only the first term to make the discussion clear. From this expression it can be deduced that the excitations which cause the sites $m$ and $n$ to be correlated over time, move as a wavepacket with a momentum distribution given by $\cos^2\alpha_k$. The SPA essentially reflects the fact that the slowest moving excitations determine the asymptotic behavior of $\delta C_{mn}$. This is because we are only considering the contribution from regions around the stationary points of $\epsilon_k$ i.e. around the point where the group velocity $d\epsilon_k/dk$ is zero. However, the distribution function $\cos^2{\alpha_k}=\chi^2\sin^2k/\epsilon_k^2$ always vanishes at $k=0,\pm\pi$. Particularly, although in the commensurate phase, the slowest moving excitations are those with momenta $k=0,\pi$, such excitations have very low amplitude and therefore do not contribute to the relaxation of the system. In other words,  because $\cos^2\alpha_k$ vanishes at the BZ boundaries, the approach to the steady state is faster ($t^{-3/2}$) in this phase. Meanwhile, for the incommensurate phase the slow excitations have a non zero density hence the decay is slower ($t^{-1/2}$).
\section{Quenching on the disorder line}\label{sec3}

\begin{figure}[t]
{\includegraphics[width=0.85\columnwidth]{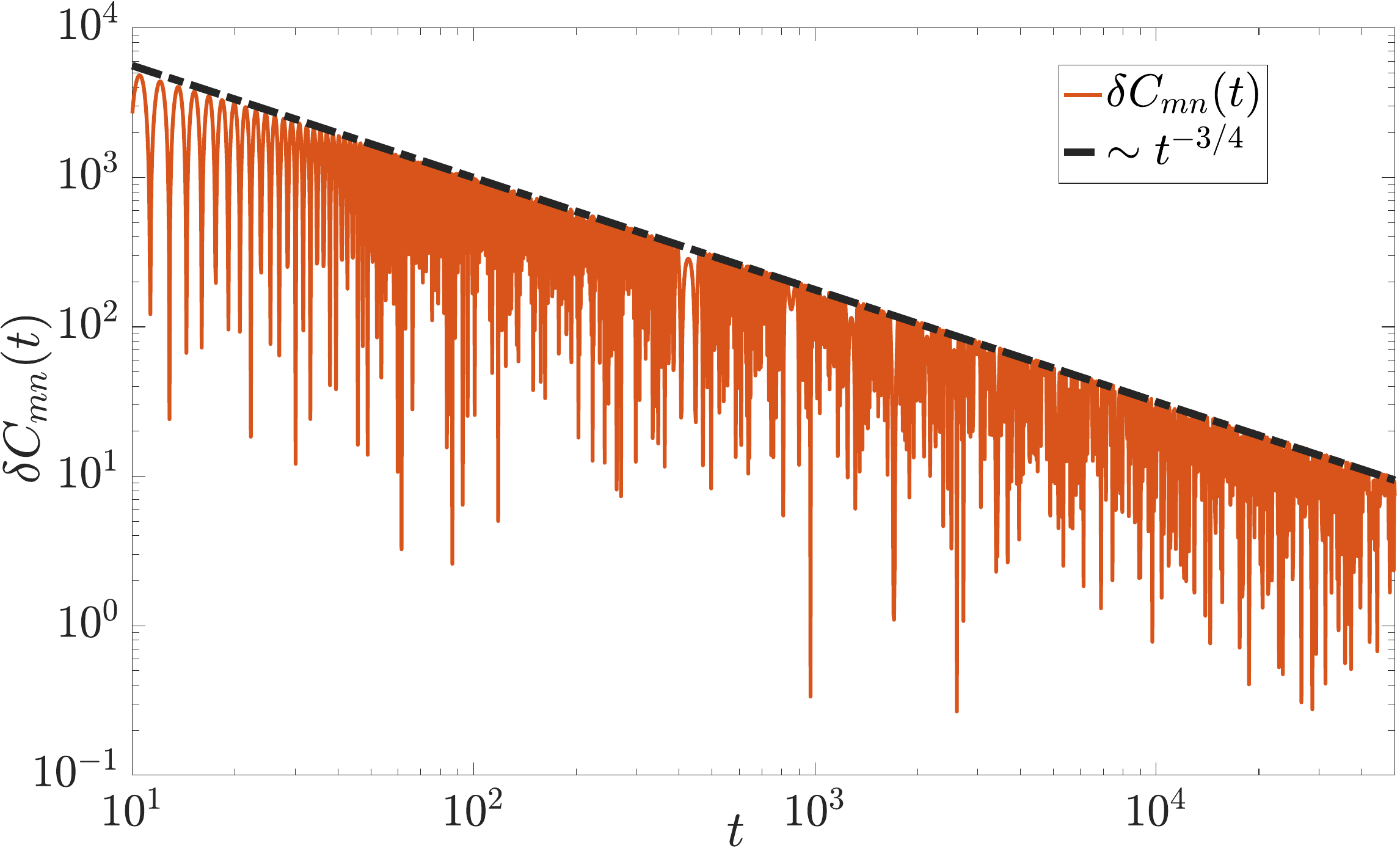}}
\caption{ \label{critical} Scaling for quench on the DL, $h=0.5$, $\chi=1/\sqrt{2}$. The scaling relation $\delta C_{mn}(t)\sim t^{-3/4}$, analytically predicted in the text is thus verified.  {The correlation functions have been scaled up by a factor of $L=200000$.}}
\end{figure}

\begin{figure}[t]
{
{\includegraphics[width=7cm, height=4cm]{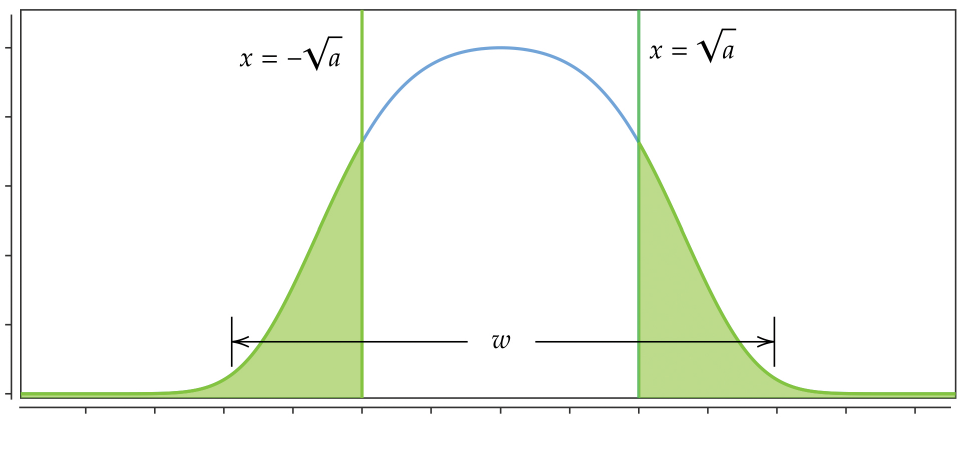}}
\caption{\label{detail}The exponential term $e^{-(ak^2+k^4)t}$ (blue) has a width $w\sim t^{-1/4}$ for $a\xrightarrow{}0$. The $k^4$ term contributes significantly in the shaded area. Crossover occurs when $w$ becomes smaller than $2\sqrt{a}$.}}
\end{figure}

\begin{figure}[t]
{\includegraphics[width=8cm, height=5cm]{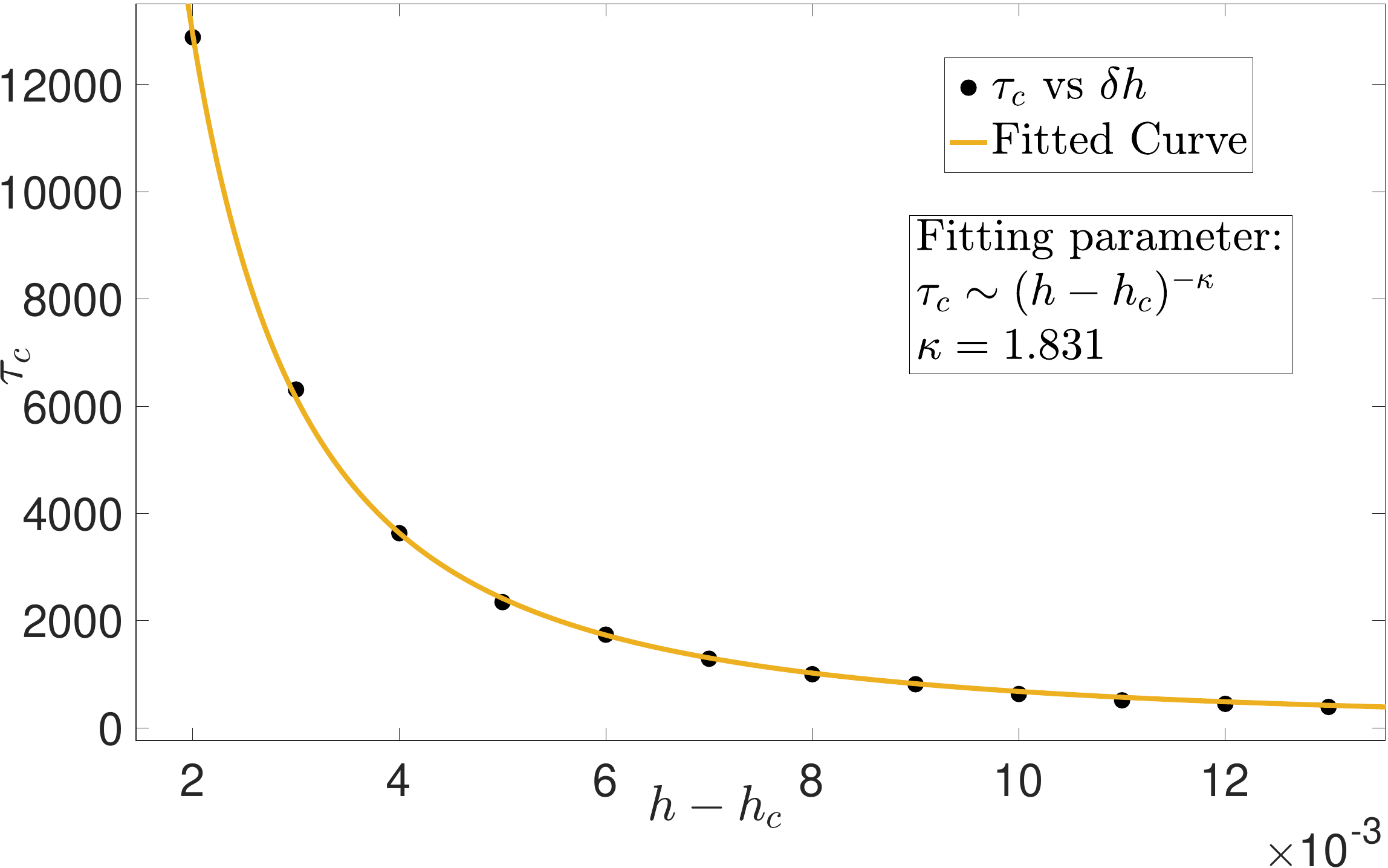}}
\caption{\label{crossoverdiv} Variation of crossover time $\tau_c$ with $h-h_c$. The numerically estimated crossover exponent matches closely with the analytic value of $\kappa=2$.}
\end{figure}

Although we have seen two different dynamical phases in Sec.~\ref{sec2}, we have not yet described how the transition occurs as we adjust the parameters and how does the system relax when the quenched Hamiltonian lies on the DL (blue line in Fig.~\ref{XYPD}). We address these issues in this section using the XY model. For a fixed $\chi$, we recall that the transition occurs at $h=h_c\equiv\sqrt{1-\chi^2}$. Consider the integral $I(t)$ again. For the SPA, we had expanded $\epsilon_k$ around $k=0$ to second order in $k$. On the DL, the second derivative of $\epsilon_k$ also vanishes, so we have to go to the fourth order contribution in $k$. The integral thus takes the form
\[
    I(t)\sim\int_{-\infty}^{\infty}dkk^2\exp\Bigg[{i\frac{d^4\epsilon_k}{dk^4}\Bigg|_{0}k^4t}\Bigg],
\]
from where we can extract the scaling,
\begin{equation}
    \delta C_{mn}(t)\sim t^{-3/4}.
\end{equation}
This scaling is verified by the numerical calculations (Fig.~\ref{critical}).
\begin{figure}[t]
{\includegraphics[width=0.85\columnwidth]{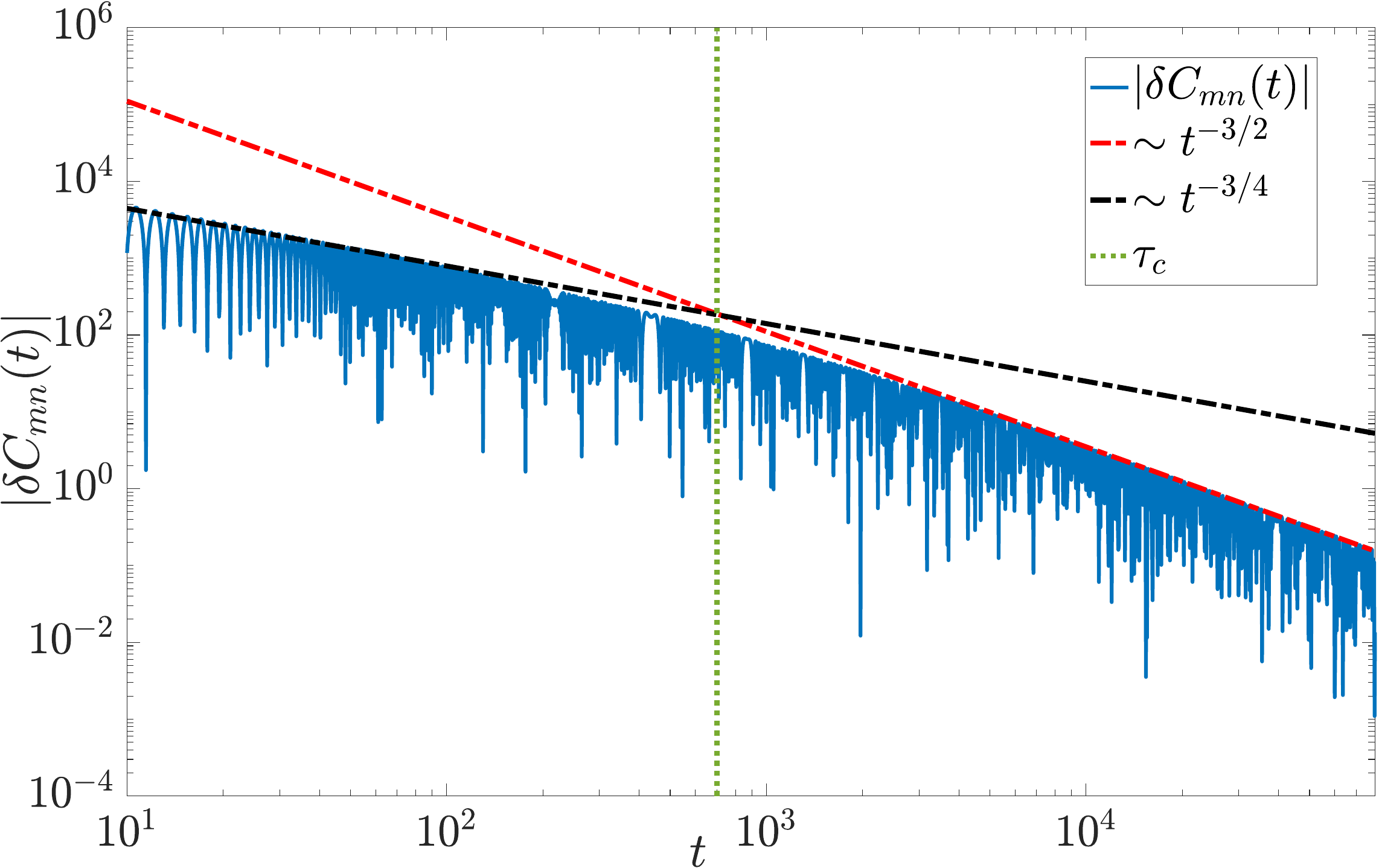}}
\caption{\label{criticalcross} Crossover from the  DL scaling ($\sim t^{-3/4}$)  to commensurate ($\sim t^{-3/2}$) scaling for a quench close to the DL ($\delta h=0.01$). The dotted green line indicates the crossover time $\tau_c$.}
\end{figure}
Near the DL, i.e.~for small values of $\delta h=h-h_c$, the second derivative of $\epsilon_k$ is still small and for intermediate times, the $k^4$ term stays significant ($t^{-3/4}$ scaling)  until a crossover to the asymptotic behavior ($t^{-1/2}$ or $t^{-3/2}$) at a time scale $\tau_c$ (Fig.~\ref{criticalcross}). 
As we approach the DL, $\tau_c$ diverges. To estimate the crossover time $\tau_c$, we look at the exponential inside the integral again: $\exp[-t(ak^2+k^4)]$, {where we have rescaled variables so that the fourth derivative of $\epsilon_k$ is equal to unity and also replaced $i$ with a minus sign for clarity. This is possible as the rotating wave approximation kills off any contribution from fast oscillating exponential. We model this decay in the contribution of fast oscillations by an exponential approximation. Particularly, it is justified to replace $i$ as it only leads to the fact that the kernel is now exponentially dying instead of rapidly oscillating and has the same effect in terms of contribution to the integral (see Appendix~\ref{appendix_1})}. The term with the coefficient $a$ is proportional to the second derivative and dictates the contribution of the {$k^4$ term in comparison to the $k^2$ term}. In particular, the $k^4$ term is important when $k^4>ak^2$ or $k>\sqrt{a}$. The $k^4$ term thus contributes to the integral in the range $\sqrt{a}<k<w$ where $w$ is the width of the exponential. The crossover occurs when the $k^4$ term stops being significantly large, that is when $\sqrt{a}\sim w$ (Fig.~\ref{detail}). When $a$ is small, $w$ is of the order of $t^{-1/4}$. Thus the crossover time obeys the relation $\sqrt{a}\sim\tau_c^{-1/4}$. The first non-zero term in the expansion of $a$ around zero is linear in $\delta h$.  {Therefore, $\tau_c$ diverges upon approaching the DL as,
\begin{equation}
    \tau_c\sim|h-h_c|^{-\kappa},
\end{equation}
with $\kappa=2$.}
Numerical estimation of $\tau_c$ and curve fitting give a value of the crossover exponent reasonably close to the analytical value (Fig.~\ref{crossoverdiv}).
We remark here, that the analysis performed above is for positive values of $\delta h$. The same analysis can be extended to the $h<h_c$ case as well, although with some subtleties. For $h<h_c$ we are in the incommensurate phase and so there is a saddle point inside the Brillouin zone as well as on the edges and center. For small $|\delta h|$, the exponential is again of the form $e^{i(ak^2+k^4)t}$ for the saddles at $k=0$ and $k=k_0$. The $k=0$ saddle has a crossover from $t^{-3/4}$ to $t^{-3/2}$ behavior as shown above. Correspondingly, the $k=k_0$ saddle has a $t^{-1/4}$ to $t^{-1/2}$ crossover for which the same argument for crossover time scaling follows. Both the crossover times scale the same way with $\delta h$, but due to the presence of multiple crossovers, it becomes difficult to numerically estimate the crossover exponent though  the analytical value of $\kappa$ remains the same. Note that these crossovers are between different scalings for the same saddle. Crossovers between different saddles are also possible, although they have a different mechanism which will be discussed in the next section. 

\section{extended Ising Model}
\label{sec:extended_ising}

The extended Ising Hamiltonian consists of a three spin interaction term loaded on top of the regular XY model \cite{sourav}. The Hamiltonian can be written as

\begin{equation}
    H_{EIM}=aH_{XY}+bH_3,
\end{equation}
where,
\begin{equation}
    H_3=-\sum_{i=1}^{L}{\sigma_i^z}\left[\left(\frac{1+\delta}{2}\right)\sigma_{i-1}^x\sigma_{i+1}^x+\left(\frac{1-\delta}{2}\right)\sigma_{i-1}^y\sigma_{i+1}^y\right].
\end{equation}
JW fermionization of $H_3$ gives
\begin{equation}
    H_3=-\sum_{i=1}^L\Big[c_{i-1}^{\dagger}c_{i+1}^{ } +\delta c_{i-1}^{\dagger}c_{i+1}^{\dagger}+\textrm{H.c.}\Big].
\end{equation}
Using similar methods as in Sec.~\ref{sec2}, we obtain the Hamiltonian matrix $\mathbf{H}_k$ in the Fourier space as $\mathbf{H}_k=\mathbf{R}_k\cdot\mathbf{\tau}$, where 
\begin{equation}
    \mathbf{R}_k = 2\begin{pmatrix}
    0,a\chi\sin{k}+b\delta\sin{2k},a(h-\cos{k})-b\cos{2k}
    \end{pmatrix}^T.
\end{equation}
The spectrum of the extended Ising model can be obtained the same way as that of the XY model. The system exhibits several phases as we tune the 5 parameters $a,b,h,\chi,\delta$. These are the same kind of phases as those described for the XY model in Fig.~\ref{XYPD}. In particular, incommensurate and commensurate phases occur for the extended Ising model as well.\\
\begin{figure}[t]
{\includegraphics[width=7.3cm, height=5.0cm]{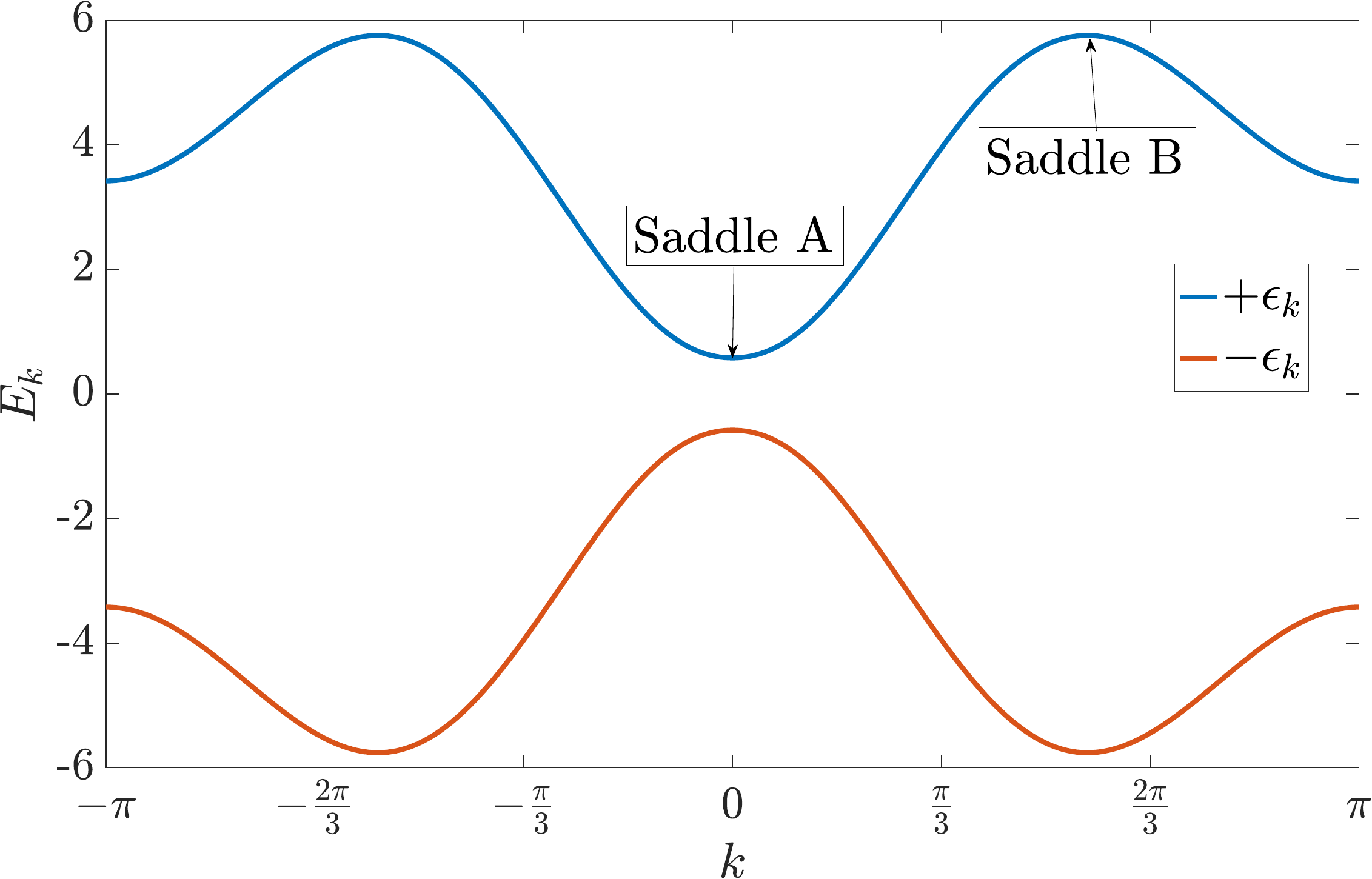}}
\caption{\label{spec3} Energy spectrum for the extended Ising model with parameter values $a=1$, $b=1$, $\chi=1/\sqrt{2}$, $\delta=0.43$ and $h=1.71$. Saddle A leads to a $t^{-3/2}$ scaling while Saddle B provides a $t^{-1/2}$ scaling.}
\end{figure}

\begin{figure}[b]
{\includegraphics[width=0.8\columnwidth]{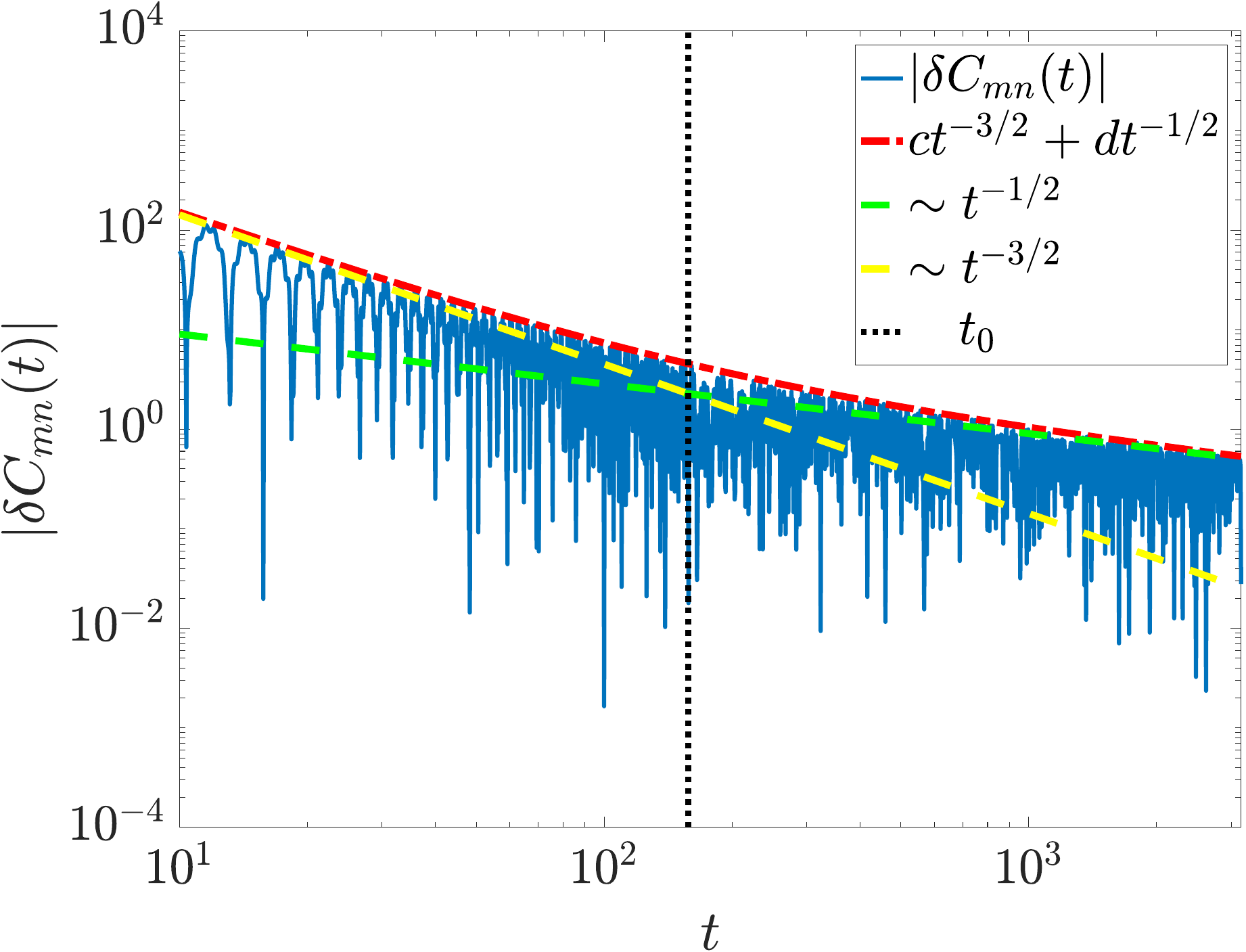}}
\caption{\label{crossover}Crossover from $t^{-3/2}$ to $t^{-1/2}$ scaling. The coefficients $c$ and $d$ have been estimated using SPA (Eqs.~\ref{IA} and \ref{IB}).  {The correlation functions have been scaled up by L=40000}. The dotted black line indicates the crossover time $t_0$.}
\end{figure}

In Sec.~\ref{sec2}, while calculating the integrals, we did not carefully include the proportionality constants in the scaling relation and the contributions from different saddle points but we obtained the correct scaling behavior nevertheless. However, in the extended Ising model, there is some non-trivial behavior due to the existence of multiple saddle points as we show. Consider the extended Ising model with parameter values $a=1$, $b=1$, $\chi=1/\sqrt{2}$ and $\delta=0.43$. Similar to previously studied quenching protocols, we start with the $h\xrightarrow{}\infty$ ground state which is the fermionic vacuum and quench the system to $h_f=1.71$ at $t=0$. The spectrum for the final Hamiltonian is shown in Fig.~\ref{spec3}. For the parameter values of (Fig.~\ref{spec3}), the spectrum is incommensurate as can be seen from the figure, so one would expect a $t^{-1/2}$ scaling as shown in Sec.~\ref{sec2}. But in this case for intermediate time scales, the scaling is closer to $t^{-3/2}$ and later asymptotes to $t^{-1/2}$ after a crossover (Fig.~\ref{crossover}). The reason for this is as follows.
To be precise, when making the SPA we should sum the integrals over all the saddle points,
\begin{equation}
    \delta C_{mn}(t)\approx\sum_{\textnormal{Saddles}}\int_{-\infty}^{\infty}dk\exp\Bigg[{i\frac{d^2\epsilon_k}{dk^2}\Bigg|_{k_0}(k-k_0)^2t}\Bigg]f(k),
\end{equation}
where $f(k)=\mathcal{A}\cos^2\alpha_k\cos{k(m-n)}$ ($\mathcal{A}$ being independent of $k$).
For the parameter values used in Fig.~\ref{spec3}, this means that $\delta C_{mn}(t)\approx I_A(t)+I_A(-t)+I_B(t)+I_B(-t)$, with

\begin{eqnarray}\label{IA}
&I_A(t)=\frac{(a\chi+2b\delta)^2e^{i2\epsilon_0t}}{8\pi[a(h-1)-b]^2}\int_{0}^{\infty} dk k^2e^{i\epsilon''_{k}(0)k^2t},\\
\label{IB}
&I_B(t)=\frac{\cos^2\beta_{k_0}\cos{k_0(m-n)}}{4\pi e^{-2i\epsilon_{k_0}t}}\int_{-\infty}^{\infty} dk e^{i\epsilon''_{k}(k_0)(k-k_0)^2t},
\end{eqnarray}
where the subscripts indicate the saddles A and B respectively, $k_0$ is the location of saddle B and $\cos\beta_k$ is defined as $\mathbf{\hat{R}}_k\cdot\mathbf{\hat{y}}$.  {The quantity $\cos\beta_k$ is similar to the quantity $\cos\alpha_k$ discussed in previous section but instead refers to the extended Ising model.} We remark that there is also the saddle at $k=\pi$ which we haven't considered, but it turns out that its contribution is small and the scaling behavior is accurately captured by the contributions of just the two saddles A and B.\\

Note that if the scaling is not in the form of a single power law but a sum of different power laws,
\[\delta C_{mn}(t)\sim a_1t^{p_1}+a_2t^{p_2}+\dots,\]
then asymptotically, the scaling relation will be dominated by the term giving the slowest decay, that is the one with the least negative power. However, at intermediate times, it may happen that because of disparity in the size of the prefactors $a_i$, a term with a power which is not the least negative, may dominate. For the case demonstrated in Figs.~\ref{spec3} and \ref{crossover}, due to the presence of two kinds of saddles, the decay is given by
\begin{equation}\label{eq:crossover}
    \delta C_{mn}(t)\sim ct^{-3/2}+dt^{-1/2}.
\end{equation}
 {Although the coefficients $c$ and $d$ are naturally complex and oscillatory in time, since we are only interested in the overall power law scaling behavior, we set all oscillatory terms and overall phase factors as unity and treat $c$ and $d$ as real, positive quantities. This is justified because the oscillatory terms do not have any interference effect as the frequencies of these oscillations are very different. This is because the frequencies correspond to the energies at different saddle points (For further details, see Appendix \ref{appendix_0}).} At smaller times (but large enough for the SPA to work) the first term dominates because $c\gg d$. At a time $t_0=c/d$ the two terms become comparable and there is a crossover from $t^{-3/2}$ to $t^{-1/2}$ behavior. The coefficients $c$ and $d$ have been calculated from Eqs.~\ref{IA} and \ref{IB} and the corresponding scaling shown in Fig.~\ref{crossover}.
\section{XY model revisited}\label{sec5}
Crossovers similar to the situations discussed in Sec.~\ref{sec:extended_ising}, are also possible for quenches in the XY model close to the DL. Recall from Sec.~\ref{sec3} that there are two different saddle points present and there is a crossover between two scalings within each saddle. However, because of reasons mentioned in the last paragraph of Sec.~\ref{sec:extended_ising}, there could be, in principle, crossovers between the two saddles. Consider the $t^{-3/2}\xrightarrow{}t^{-1/2}$ crossover: the $k=0$ saddle integral is of the form
\[
I\sim\int dk k^2 e^{i(ak^2+k^4)t},
\]
while the $k=k_0$ integral is of the form
\[
I'\sim\int dk \sin^2k_0 e^{i(ak^2+k^4)t}\approx k_0^2\int dk e^{i(ak^2+k^4)t},
\]
where we have used the fact that $k_0$ is small near the disorder line. Also, since $k_0\sim \delta h$ for small $k_0$ by definition of $k_0$, the crossover time $t_0$ scales as
\begin{equation}
\label{scal}
    t_0\sim \frac{c}{d}\sim |h-h_c|^{-2}.
\end{equation}
Note that by similar reasoning, $t^{-3/4}\xrightarrow{}t^{-1/4}$ and $t^{-3/2}\xrightarrow{}t^{-1/4}$ crossovers are also possible with crossover times diverging as $\delta h^{-1}$ and $\delta h^{-3}$ respectively. We have actually observed the first of these in simulations, but not the second kind. The reason for this is that near the disorder line, the $t^{-3/2}\xrightarrow{}t^{-1/4}$ crossover time diverges faster than all other crossover times, so that both the terms $t^{-3/2}$ and $t^{-1/4}$ are already killed by the time the $t^{-3/2}\xrightarrow{}t^{-1/4}$ crossover takes place.\\

To summarise, in this section and at the end of Sec.~\ref{sec:extended_ising}, we have discussed the dynamical crossover of relaxation dictated by a saddle which corresponds to a $t^{-3/2}$ scaling to a saddle which corresponds to a $t^{-1/2}$ scaling for both the extended Ising and the XY model. We emphasize here that this crossover is between the scalings obtained following quenches into the commensurate ($t^{-3/2}$) and incommensurate ($t^{-1/2}$) phases. This crossover is conceptually different from that discussed in Sec.~\ref{sec3} which is between the scalings following quenches on the disorder line ($t^{-3/4}$ scaling) and into the commensurate/incommensurate phases ($t^{-3/2}$/$t^{-1/2}$ scalings). It is the latter kind of crossover for which we can define the crossover exponent $\kappa$ and it indeed equals two for approaches to the boundary from either side. The exponent two appearing in Eq.~\ref{scal} is specific to this special situation   and the exponent for such a crossover may not even be always defined.

\section{Robustness against integrability breaking perturbations}\label{sec6}

We now probe the robustness of the relaxation crossover behavior against integrability breaking perturbations. To break the integrability of the XY system (Eq.~\ref{XYHam}), we introduce anisotropic ferromagnetic next nearest neighbour interactions, thereby modifying the Hamiltonian to
\begin{equation}
H^{\prime}=H_{XY}-JV,
\end{equation}
such that
\begin{equation}
	V=\sum\limits_i\sigma_i^x\sigma_{i+2}^x.
\end{equation}
Here, the coupling $J>0$ controls the strength of the integrability breaking interactions. Numerical procedures applied in the previous sections, particularly the decomposition into decoupled single-particle momentum sectors are no longer effective in the nonintegrable scenario. We therefore resort to brute-force exact diagonalization \cite{quspin17,quspin19} (ED) to probe the dynamical relaxation behaviors following a quench. For numerical advantage, we consider translationally invariant local operators which are linear in the two point fermionic correlators to study the relaxation towards a steady state. Two examples of such observables are,
\begin{equation}
	C_1=\frac{1}{L}\sum_i\sigma_i^z,
\end{equation}
and
\begin{equation}
	C_2=\frac{1}{L}\sum\limits_i\sigma_i^x\sigma_{i+1}^x.
	\label{bilinear}
\end{equation}

\begin{figure*}
\subfigure[]{
	\includegraphics[width=7cm,height=4.5cm]{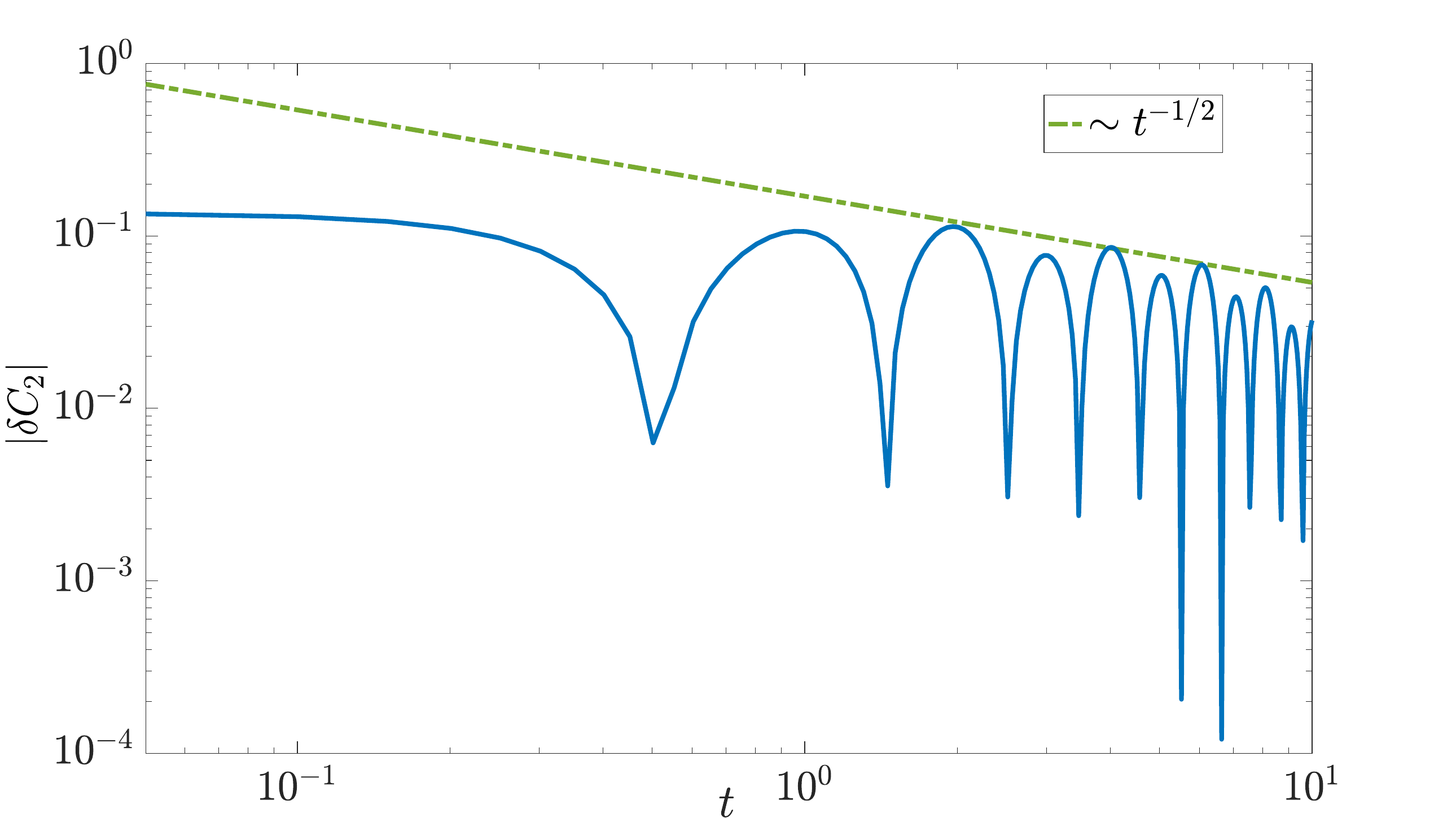}
	\label{ED1}}
\hspace{1.0cm}
\subfigure[]{
	\includegraphics[width=7cm,height=4.5cm]{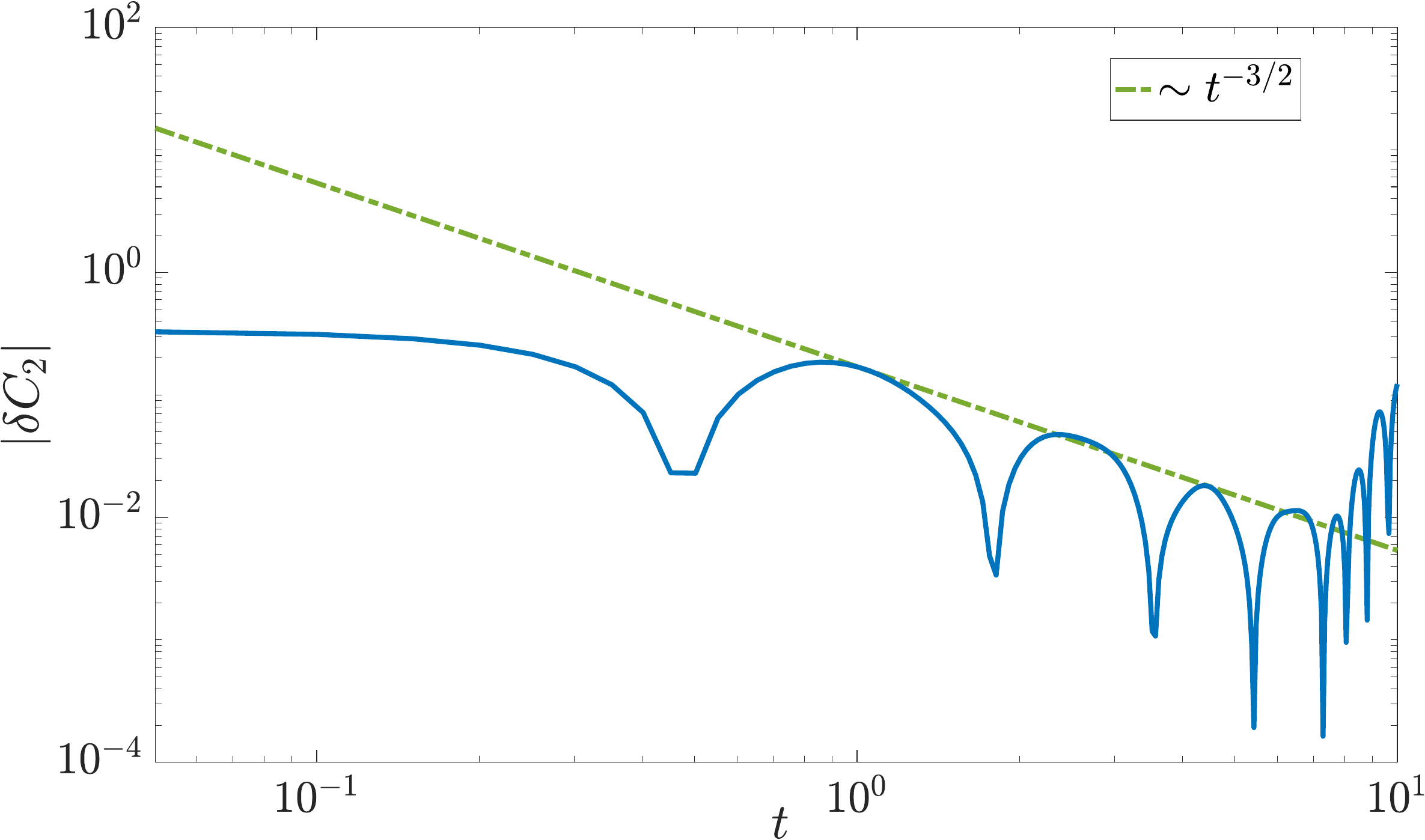}
	\label{ED2}}
\caption{Decay of bilinear spin correlations $C_2$ defined in Eq. \eqref{bilinear} under weak integrability breaking ($J=0.1$) for parameters (a) $h=0.25$, $\chi=\frac{1}{\sqrt{2}}$, such that $H_{XY}$ is in the incommensurate phase.
 (b) The correlations for parameters $h=0.75$, $\chi=\frac{1}{\sqrt{2}}$, such that $H_{XY}$ is in the commensurate phase. The integrable scalings ($\sim t^{-1/2}$ in
case (a) and $\sim t^{-3/2}$ in the case (b)) survive until around $t=J^{-1}=10$ where the finite size effects take over.
  {The system size is $L=26$ in both cases.}}	
 \label{ED_weak}
\end{figure*}

Specifically, we look at the temporal distance of the local operator $C_2$ from its asymptotic steady state value $C_2(\infty)$ by studying the quantity
\begin{equation}
	|\delta C_2|=\left|C_2(t)-C_2(\infty)\right|.
\end{equation}
Since the final Hamiltonian $H^{\prime}$ generating the time evolution is also 
translationally invariant, the evolution can be decoupled into independent many-body momentum eigensectors labeled by $K$. As in the integrable situation, we start from an unentangled polarised initial state $\ket{\psi(0)}=\ket{\uparrow\uparrow...}$, which essentially lies in the $K=0$ momentum sector. This suggests that the dynamics is confined to only one momentum sector, i.e., $K=0$. Thus, it is  sufficient to probe the dynamics and thermalization of the local operators $C_1$ and $C_2$ in this subspace, with a significantly reduced Hilbert space dimension.  In Fig.~\ref{ED1}-\ref{ED2}, we observe that the crossover between different power-law relaxations indeed survive under quenches breaking integrability weakly. In finite size systems, it is straightforward to estimate the strength of the nonintegrable perturbation neccessary to significantly affect the relaxation dynamics and the subsequent destruction of slow power-law decays. The dynamics of local observables following a quench in finite size systems is generically known to show recurrences after time scales $t_r$ of the order of the system size $L$ \cite{Essler_2016}. At the same time, the integrability breaking perturbation $J$ sets a time scale, $t_c\sim J^{-1}$: for $t\gtrsim t_c$, the perturbation prominently dominates the dynamics. Consequently, the slow relaxation crossovers start to vanish as chaotic dynamics take over for $t\gtrsim t_c$. Hence, in finite size systems, for the dynamical transitions observed following integrable quenches (see Appendix~\ref{appendix_2}) are expected to survive only when the integrability breaking is sufficiently small, i.e., $J\lesssim L^{-1}$.\\

\begin{figure*}
\subfigure[]{
	\includegraphics[width=6.5cm,height=4.0cm]{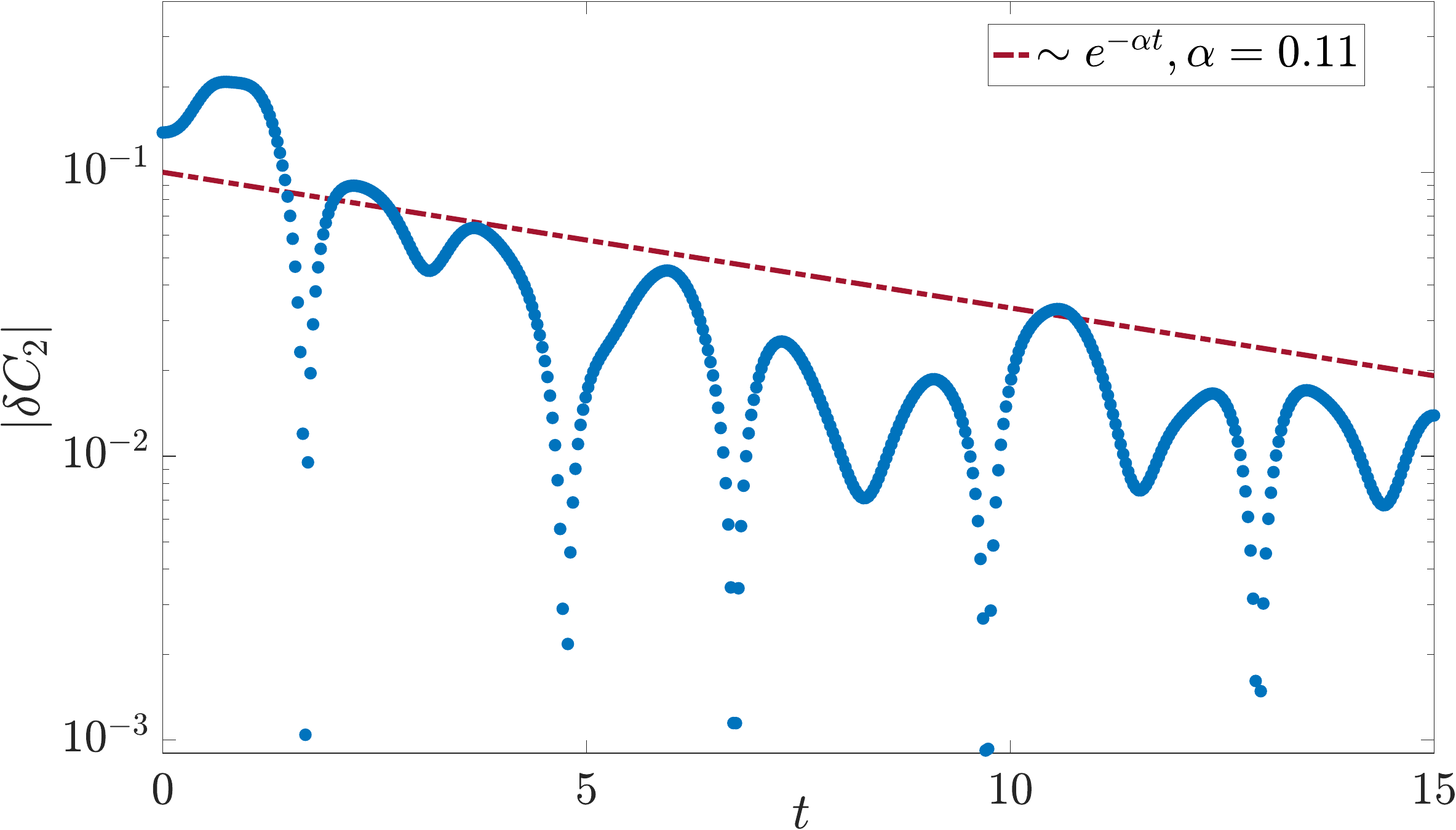}
	\label{EDs1}}
\hspace{1.5cm}
\subfigure[]{
	\includegraphics[width=6.9cm,height=4.3cm]{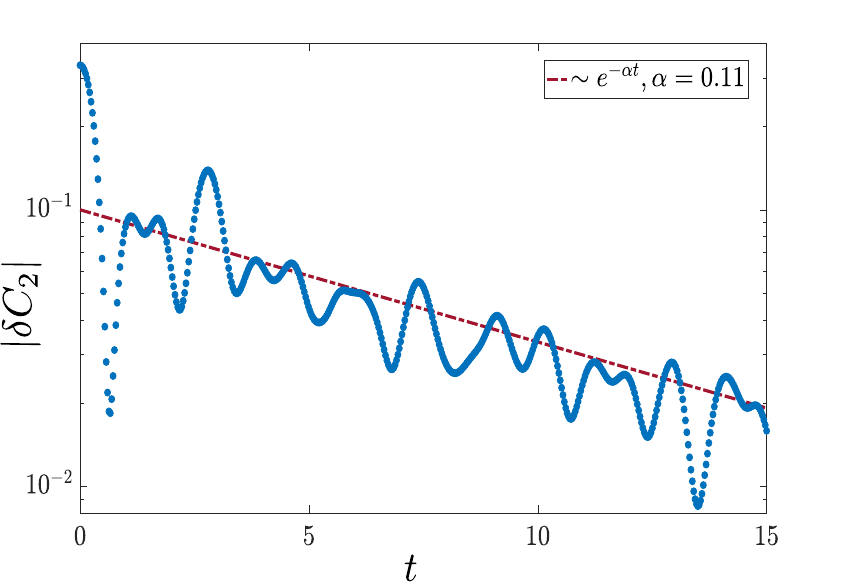}
	\label{EDs2}}
\caption{Decay of bilinear spin correlations $C_2$ defined in Eq. \eqref{bilinear} following a strong chaotic quench with the non-integrability strength $J=1$ for parameters (a) $h=0.25$, $\chi=\frac{1}{\sqrt{2}}$ and (b) $h=0.75$, $\chi=\frac{1}{\sqrt{2}}$. Unlike results presented in Fig.~\ref{ED_weak}, the integrable scalings do not survive in this case. Instead an
approximate exponential decay apparently emerges for such quenches.  {The system size is $L=26$ in both cases.}}		
\end{figure*}

On the contrary, under strongly chaotic quenches, i.e., when the integrability breaking perturbation becomes comparable in strength with the natural time scales of the system at $J\sim\mathcal{O}(1)$, all slow relaxations and the crossover behavior observed for integrable quenches are washed out. In Fig.~\ref{EDs1}-\ref{EDs2}, we demonstrate that the integrable scalings do not hold true in this case even with the same quenching parameters. Rather, the relaxation to a steady state is found to be exponentially fast \cite{alba19,lin17}. However, due to numerical limitations, we did not observe a crossover from the slow power law relaxation to an exponential decay of correlations. Although we present the analysis using the operator $C_2$, the magnetisation density $C_1$ has also been observed to show qualitatively similar relaxation behaviors. 

\section{Conclusion and future directions}
\label{sec_conclusion}
By studying the relaxation of quenched quantum many body systems, we have demonstrated the class of dynamical phase transitions, first reported in Ref.~\cite{sen2016} for periodically driven systems. These transitions are described in terms of the relaxation of correlations to a steady state after the quench, specifically, at late times the difference of the correlations from their steady state value decays as a power law in time.  The transition is reflected as the change of the exponent of this power law relaxation behavior. Thus, following integrable quenches, we observe slow/fast power law decays depending on the single-particle spectrum of the quenched Hamiltonian. {Particularly, we have connected this transition to the momentum distribution and group velocity of the slowest moving excitations in the system following the quench. Interestingly, this connection might allow one to construct various Hamiltonians by working backwards from arbitrarily chosen momentum distribution of quasi-particles. Studying such constructions in the future may lead to a deeper understanding of the origin of such dynamical transitions.}\\

Additionally, in the integrable XY model, we have found the scaling of decay of correlations for quenches on the disorder line to be different than that in either of the phases. The approach of the quenched Hamiltonian to the disorder line is also characterized by a diverging time scale. This time scale diverges as a power law of the quenched transverse magnetic field as it approaches the disorder line. Whether this crossover exponent reflects an underlying universality is an interesting direction of further research. {However, we note that a diverging time scale implies a vanishing energy scale in the system, purely on dimensional grounds. Moreover, the maximum group velocity of excitations provides us with a characteristic velocity scale for the system. Combining this velocity scale with the time scale gives us a length scale which diverges at the boundary of the transition. For systems at equilibrium, a vanishing energy scale and a diverging length scale are often indicative of a quantum critical point \cite{sachdev_2011,ADbook}. This gives us a hint that it could be worthwhile to investigate the relation of the dynamical phase transitions described in this work with criticality and universality.}\\

{Besides the distinct power-law decays, we have also found that under certain conditions, there could occur a different kind of relaxation crossover. In such a situation, the scaling of the correlation decay is different from the expected asymptotic scaling until a crossover time $t_0$. Although one may argue that this phenomenon is trivial since the asymptotic scaling is expected only at large times (in thermodynamically large systems), this behavior could have some non-trivial implications for experimental setups. This is because of finite size effects where the excitations cannot keep moving ad infinitum. For example, under open boundary conditions, upon collision with the boundary, the excitations reflect back within the system and cause revival effects \cite{Essler_2016}. If the crossover time is larger than the time taken for excitations to move from one end of the system to the other, the asymptotic scaling would never be observed.} In addition to integrable quenches, we also probe the robustness of the dynamical transitions in the case of nonintegrable quenches. Interestingly, the crossover relaxation behavior is seen to persist even under weakly chaotic quenches. Particularly, in finite size systems of length $L$, it is possible to identify a finite non-zero energy scale ($\sim L^{-1}$) such that sufficiently small integrability breaking perturbations in the quenched Hamiltonian do not affect the integrable ralaxation behaviors. However, following generic quenches in strongly chaotic systems, the local observables are observed to relax exponentially fast.\\

{The advent of experimental setups consisting of lattices of ultracold atoms or trapped ions has made it possible to verify theoretical predictions in lattice models experimentally \cite{Zhang2017}. In an experimental setting, one could make measurements of the transverse magnetization, $C_1$. This should give the scaling behaviors we have seen numerically and analytically as $\sigma_i^z=1-2c^{\dagger}_ic_i$ is linearly related to the same site correlators. According to our simulations, a system of size $\sim L=40$ should be adequate to see the dynamical phase transitions. Smaller sizes may lead to finite size effects such as revivals kicking in before the asymptotic scaling behavior is attained. Particularly, Ref.~\cite{Zhang2017}, has already measured late time bilenear spin correlations in one dimensional spin systems simulated with 53 trapped ions. Hence, our predictions are well within the reach of state of the art experimental setups.}\\

\textit{Note added:} While this manuscript was in preparation
we came to know about a similar work by Aditya, Samanta, Sen, Sengupta and Sen \cite{aditya2021dynamical}, predicting anomalous power laws in periodically driven systems. Our results agree wherever a comparison is possible.

\section*{Acknowledgements}
A.A.M. acknowledges support from IIT Kanpur and an INSPIRE fellowship, DST, India. S.B. acknowledges support from a PMRF fellowship, MHRD, India.  A.D. acknowledges financial support from SPARC program, MHRD, India and SERB, DST, New Delhi, India. S. M. acknowledges SPARC Program, DST and Arijit
Kundu for support. We acknowledge Sourav Bhattacharjee, Madhumita Sarkar, Arnab Sen, Diptiman Sen and Krishnendu Sengupta for comments. 

\appendix
\section{Coefficients for power law scalings}
\label{appendix_0}
\begin{figure}[ht]
	{\includegraphics[width=0.85\columnwidth,height=0.55\columnwidth]{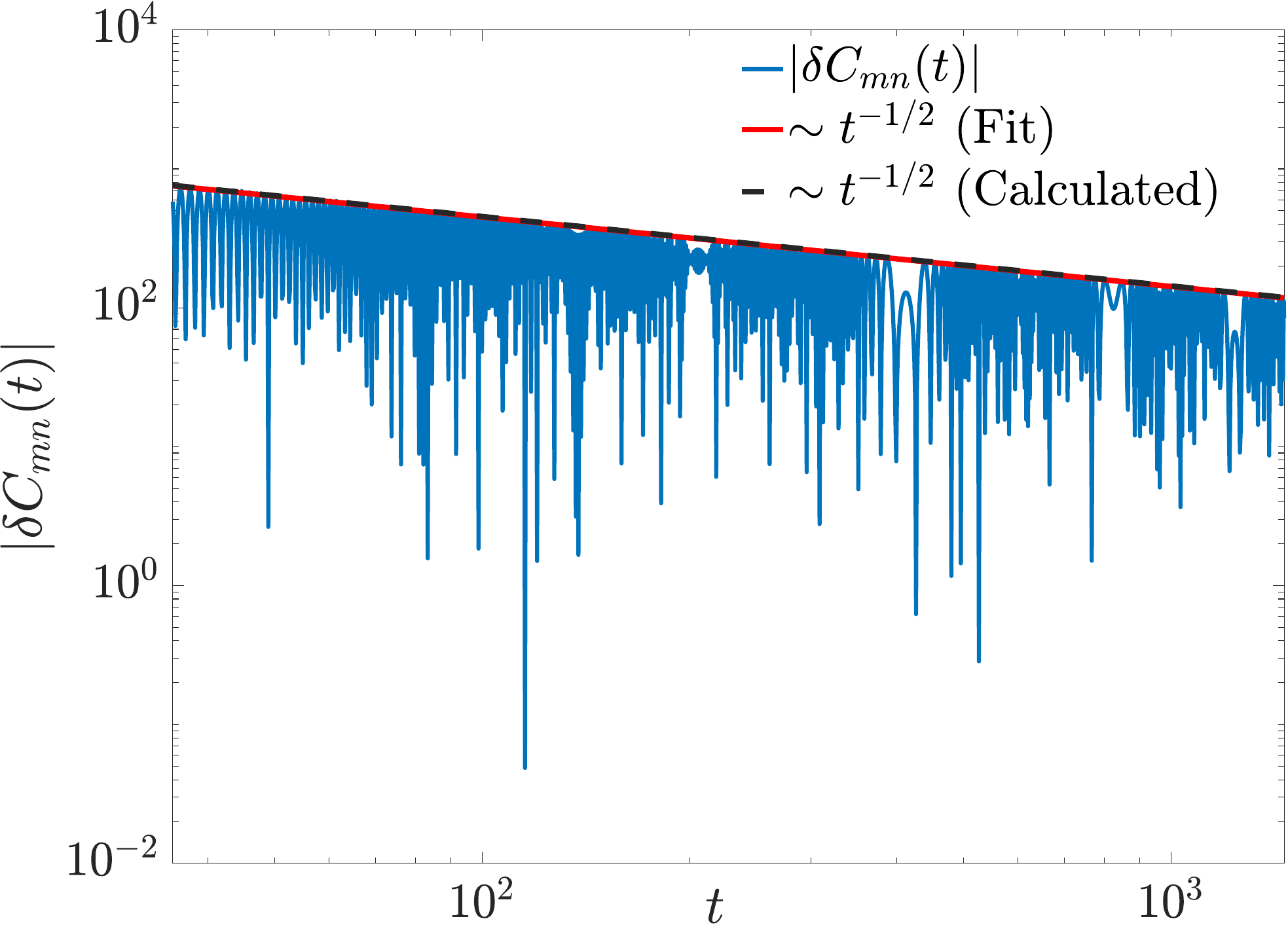}}
	\caption{\label{fig:appfit} {Comparison between scaling envelopes obtained from fitting (solid red line) and exact coefficients (broken black line) for a quench to the incommensurate phase (see also Fig.~\ref{incomm}). It can be clearly seen that the two curves match very closely.} }
\end{figure}

 {In order to calculate the coefficients for the various power law scalings, recall from Sec.~\ref{sec2} that the difference of the correlation function from its steady state value, $\delta C_{mn}(t)$ may be written as a sum of two integrals, $\delta C_{mn}(t)=I(t)+I(-t)$. For a quench to the incommensurate phase, for large $t$, the SPA gives (Eq.~\ref{eq:SPAincomm}) 
\[I(t)\approx \mathcal{C}e^{i2\epsilon_{k_0}t}\int_{-\infty}^{\infty}dk\exp\Bigg[{i\frac{d^2\epsilon_k}{dk^2}\Bigg|_{k_0}(k-k_0)^2t}\Bigg],\]
with $\mathcal{C}=\cos^2\alpha_{k_0}\cos{[k_0(m-n)]}/4\pi$. The integral is now in the Gaussian form and performing the integration gives
\[I(t)=\mathcal{C}e^{i(2\epsilon_{k_0}t+\frac{\pi}{4})}\sqrt{\frac{\pi}{\epsilon''_{k_0}}}t^{-1/2},\]
where $\epsilon''_{k_0}=\frac{d^2\epsilon_k}{dk^2}|_{k_0}$. Thus,
\[\delta C_{mn}(t)=2\mathcal{C}\sqrt{\frac{\pi}{\epsilon''_{k_0}}}\cos{\left(2\epsilon_{k_0}t+\frac{\pi}{4}\right)}t^{-1/2}.\]
Now, we are only interested in the overall scaling of the correlations, so we may set the oscillating cosine term to unity for the scaling relation. This gives
\[\delta C_{mn}(t)\sim2\mathcal{C}\sqrt{\frac{\pi}{\epsilon''_{k_0}}}t^{-1/2}.\]
For a quench to the commensurate phase, the integral $I(t)$ (under SPA) is given by 
\[I(t)\approx \frac{1}{4\pi}e^{i2\epsilon_0t}\int_{-\infty}^{\infty}dk\cos^2{\alpha_k}\exp\Bigg[{i\frac{d^2\epsilon_k}{dk^2}\Bigg|_0k^2t}\Bigg].\]
Now, the support of the exponential kernel is around $k=0$. Therefore, we may expand $\cos{\alpha_k}=2\chi\sin{k}/\epsilon_k$ around $k=0$. Using $\sin{k}\approx k$ for small $k$, we get
\[\cos^2{\alpha_k}\approx \frac{4\chi^2k^2}{\epsilon_0^2}.\]
Thus, $I(t)$ becomes
\[I(t)\approx \frac{\chi^2}{\pi\epsilon_0^2}e^{i2\epsilon_0t}\int_{0}^{\infty}dkk^2\exp\Bigg[{i\frac{d^2\epsilon_k}{dk^2}\Bigg|_0k^2t}\Bigg]\]
\[=\frac{\chi^2}{4\pi\epsilon_0^2}e^{i(2\epsilon_0t+\frac{3\pi}{4})}\sqrt{\frac{\pi}{\epsilon_0''^3}}t^{-3/2},\]
where  $\epsilon''_{0}=\frac{d^2\epsilon_k}{dk^2}|_{0}$. Note that the lower limit of integration is $0$ and not $\infty$. This is because the saddle is located at the Brillouin zone edge and so only one half of the saddle contributes. We also remark here that the saddle at $k=\pi$ has a very small contribution. This is because the contribution is proportional to $1/\epsilon_k^2$ and the energy at $k=\pi$ is significantly greater than the energy at $k=0$ (see Fig.~\ref{f1}). Therefore, we may neglect this other saddle. Thus,
\[\delta C_{mn}(t)=\frac{\chi^2}{2\pi\epsilon_0^2}\cos{\left(2\epsilon_0t+\frac{3\pi}{4}\right)}\sqrt{\frac{\pi}{\epsilon_0''^3}}t^{-3/2}.\]
For similar reasons as above, the scaling is given by 
\[\delta C_{mn}(t)\sim\frac{\chi^2}{2\pi\epsilon_0^2}\sqrt{\frac{\pi}{\epsilon_0''^3}}t^{-3/2}.\]
Now, for quenches to the disorder line, $\epsilon_0''=0$ and due to the symmetry of $\epsilon_{k}$ around $k=0$, $\epsilon_0^{(3)}=0$. This means that we need to go to the fourth order in $k$ in the expansion of $\epsilon_k$. Moreover, the same argument for the expansion of $\cos^2\alpha_k$ holds. Thus, the SPA for the integral $I(t)$ gives
\[I(t)\approx \frac{\chi^2}{\pi\epsilon_0^2}e^{i2\epsilon_0t}\int_{0}^{\infty}dkk^2\exp\Bigg[{i\frac{d^4\epsilon_k}{dk^4}\Bigg|_0\frac{k^4t}{12}}\Bigg]\] 
\[=\frac{\chi^2}{2\pi\epsilon_0^2}e^{i(2\epsilon_0t+\frac{3\pi}{8})}\left(\frac{108}{{\epsilon_0''''}^3}\right)^{1/4}\Gamma\left(\frac{3}{4}\right)t^{-3/4}.\]
Going through the same process as above cases, we get the scaling relation,
\[\delta C_{mn}(t)\sim \frac{\chi^2}{\pi\epsilon_0^2}\left(\frac{108}{{\epsilon_0''''}^3}\right)^{1/4}\Gamma\left(\frac{3}{4}\right)t^{-3/4}.\]
For the extended Ising model, the coefficients $c$ and $d$ appearing in Eq.~\ref{eq:crossover} may be calculated in a similar way as the coefficients for the $t^{-3/2}$ and $t^{-1/2}$ scalings respectively. This gives
\[c=\frac{\cos^2{\beta_{k_0}}\cos{[k_0(m-n)]}}{2\pi}\sqrt{\frac{\pi}{\epsilon_{k_0}''}},\]
\[d=\frac{(a\chi+2b\delta)^2}{2\pi\epsilon_0^2}\sqrt{\frac{\pi}{\epsilon_0''^3}},\]
neglecting all oscillatory behaviors. We present in Fig.~\ref{fig:appfit} a comparison between the curves obtained from exact coefficients and fitting for the case of a quench to the incommensurate phase. It can be seen that the two curves closely match. The remaining cases match similarly.}

\section{Approximating fast oscillating exponentials}
\label{appendix_1}
\begin{figure}[ht]
	{\includegraphics[width=0.85\columnwidth,height=0.55\columnwidth]{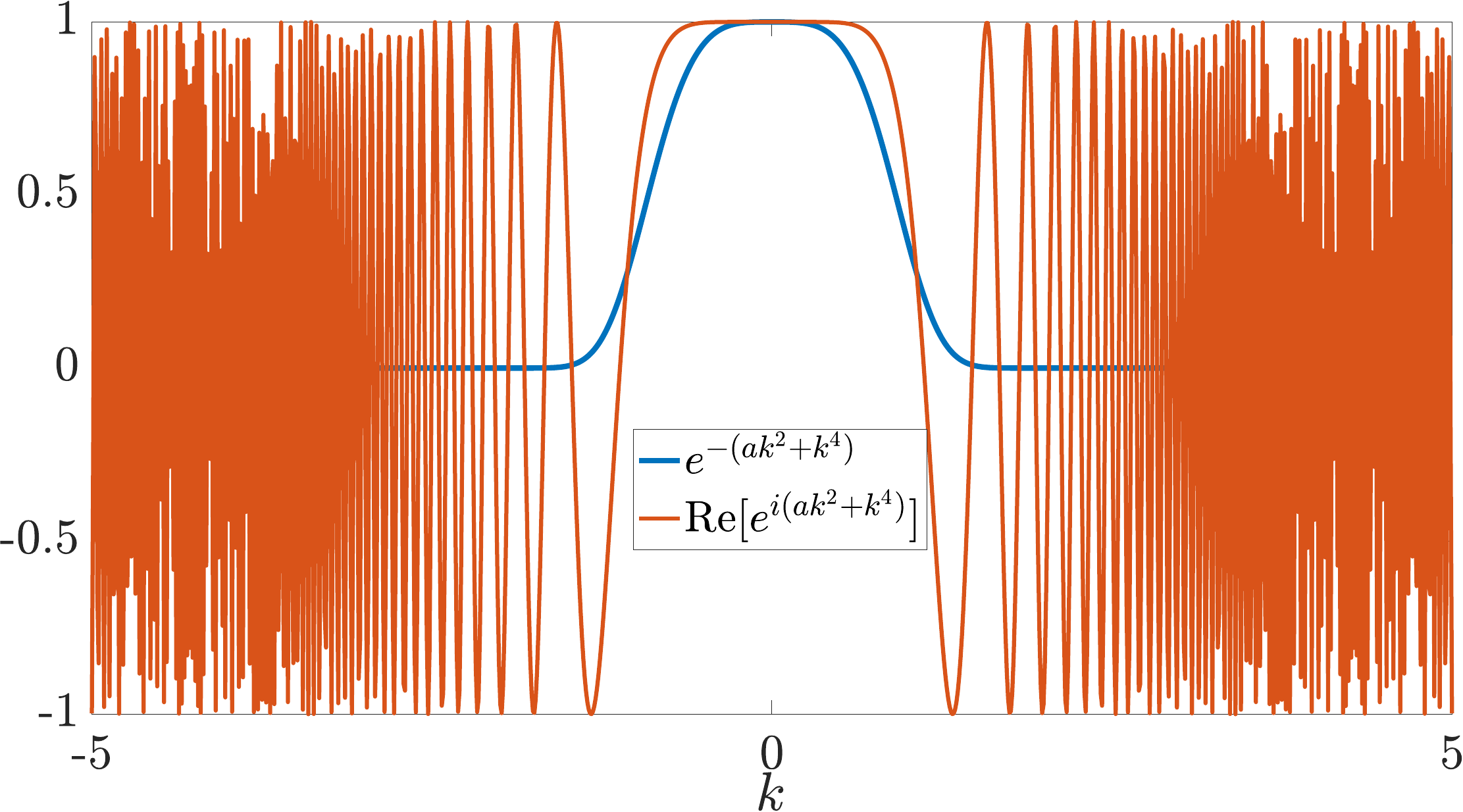}}
	\caption{\label{exp_app}The kernel $e^{i(ak^2+k^4)}$ can be replaced by $e^{-(ak^2+k^4)}$ as the contribution to the integral from $e^{i(ak^2+k^4)}$ is significant only within the width of $e^{-(ak^2+k^4)}$. Beyond this width, the rapid oscillations cause the contribution to go to zero.}
\end{figure}
In Sec.~\ref{sec3} of the main text, while discussing the divergence of crossover time, we replaced the factor of $i$ in the exponential term with a minus sign in order to keep the discussion clear. We justify the choice here. Note first that the argument in Sec..~\ref{sec3} was dependent only on the width of the exponential kernel and not on exact numerical values. The kernel contributes as part of an integral only within some width $w$ around its stationary point. Beyond this width $w$, the contributions become negligible. This is either due to exponential suppression as in the case of a kernel like $e^{-f(x)}$ with $f(x)$ real and positive or due to rapid oscillations cancelling out the contributions as in the case of a kernel like $e^{if(x)}$ with real $f(x)$. In particular, for our case where the kernel is of the second kind, the width remains the same when the $i$ is replaced by a negative sign (see Fig.~\ref{exp_app}). \\

We remark here that although replacing the oscillating exponential with a decaying one gives the correct scaling behavior, it leaves out  oscillatory behavior of
the correlators which we have not focused on in this paper. However,  the oscillatory nature is also seen to exhibit non-trivial behaviours. For a detailed discussion of this phenomenon, see Ref.~\cite{aditya2021dynamical}.

\section{Exact diagonalization results for integrable case}
\label{appendix_2}
\begin{figure}
	\subfigure[]{
		\includegraphics[width=7.0cm,height=5.0cm]{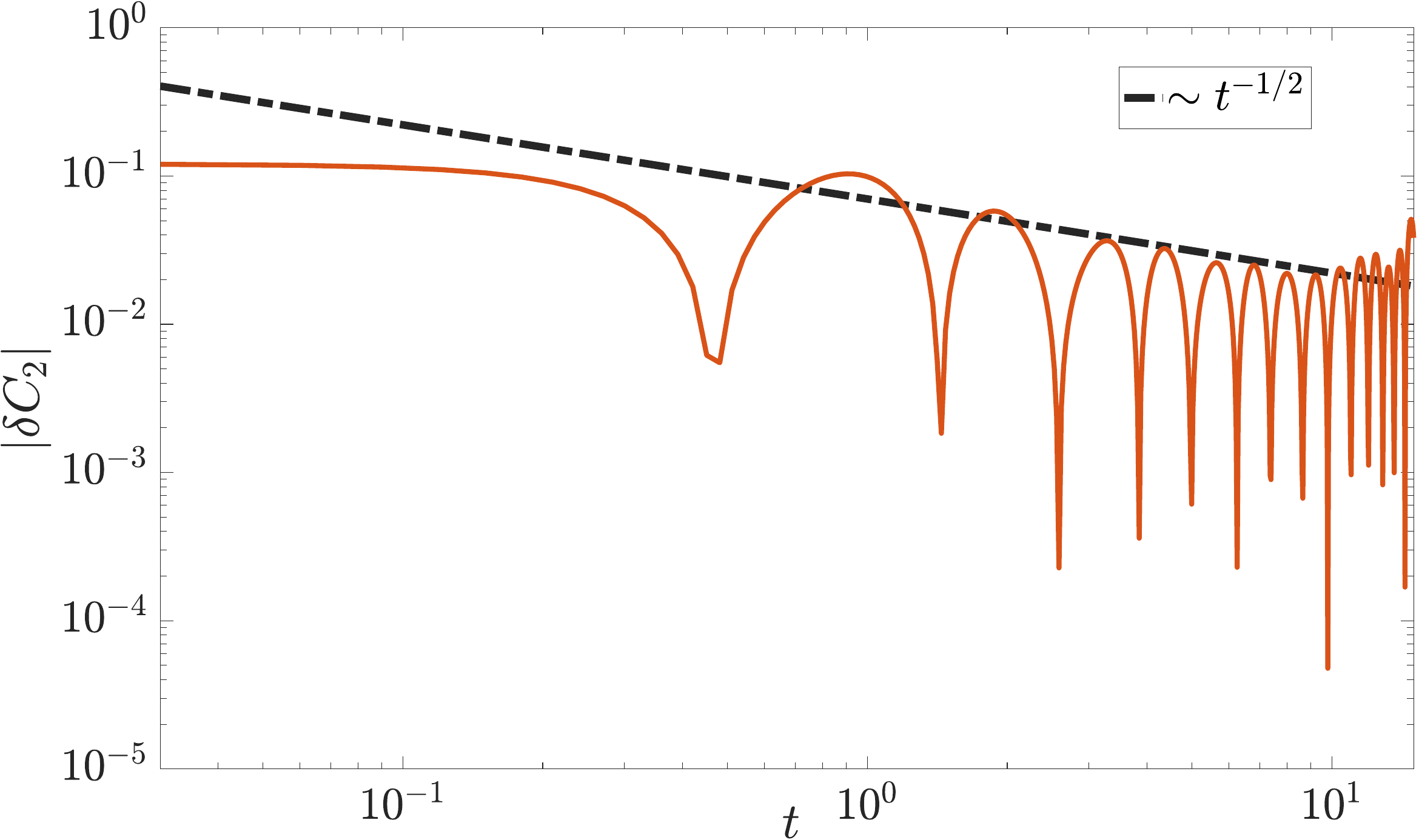}
		\label{EDint1}}
	\hspace{1.5cm}
	\subfigure[]{
		\includegraphics[width=7.0cm,height=5.0cm]{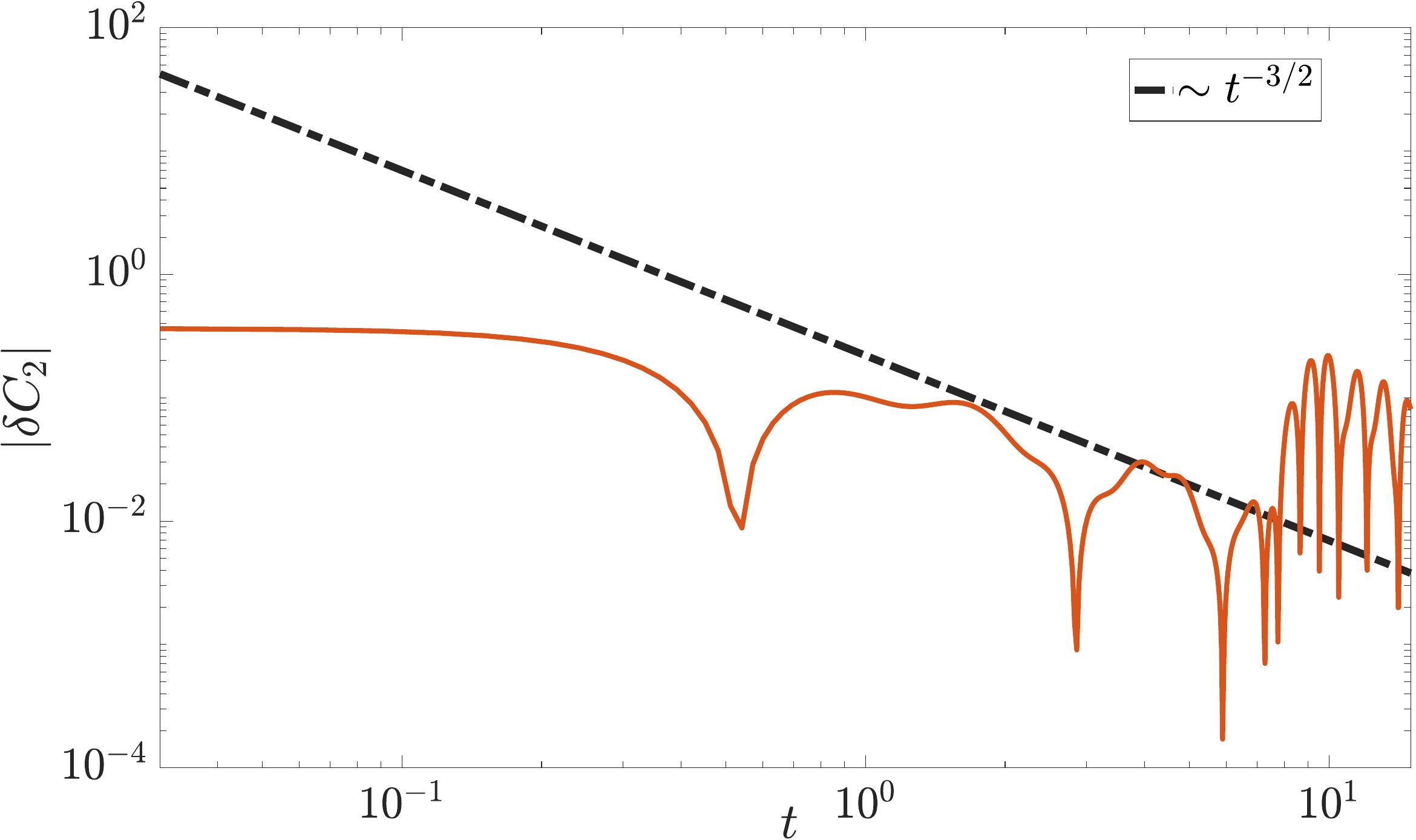}
		\label{EDint2}}
	\caption{ED results for integrable system. (a) We find a $t^{-1/2}$ scaling  following a quench to the incommensurate phase with parameters $h=0.25$, $\chi=\frac{1}{\sqrt{2}}$.  (b) On the other hand,  $t^{-3/2}$ scaling is observed following a quench to the commensurate phase with  $h=0.75$, $\chi=\frac{1}{\sqrt{2}}$.  {The system size $L=26$ in both cases.} }	
\end{figure}
The scaling relations were obtained in Sec..~\ref{sec2} through a decomposition of the Hamiltonian $H_{XY}$ into $2\times2$ single particle momentum sectors. Similar results were obtained through an exact diagonalization procedure applied on the full Hamiltonian with some caveats due to the small system size (L=26).  {Note that as both the Hamiltonian and the probe operators have been taken to be translationally invariant, all the results are obtained in only one momentum block $K=0$, which hosts the completely polarised initial state $\ket{\uparrow\uparrow...}$. Because of the translational invariance and block decomposition into decoupled momentum sectors, the effective basis dimension reduces significantly. Furthermore, we do not need to solve for the complete spectrum, as finding only the ground state is sufficient to construct the initial state of the system for generic quenching schemes. Following that, we numerically solve the time dependent Schrodinger equation to simulate the quench.}

For a system of this size, it takes a recurrence time $t_r\sim10$ for the excitations to travel around the system (for periodic boundary conditions) or to reflect from the edges of the system (for open boundary conditions). Beyond this time scale, the scalings are not expected to hold. In Figures~\ref{EDint1} and \ref{EDint2}, it can be seen that the respective scalings hold true until a time $t\approx10$. Beyond this, time dependent part of the correlators suddenly rises and the whole process recurs on the time scale $t_r$.\\

\pagebreak

\bibliography{bibliography}

\end{document}